\documentclass[useAMS,usenatbib]{mn2e}
\usepackage{graphicx,amssymb,amsmath}
\usepackage{mnras_macros}
\usepackage[dvips]{color}

\newcommand{\dm}{{\rm d}}

\title[The origin of pseudobulges in cosmological simulations]
{The origin of pseudobulges in cosmological simulations of galaxy formation}
\author[T. Okamoto]{Takashi
Okamoto$^{1}$\thanks{E-mail: 
tokamoto@ccs.tsukuba.ac.jp}\\ 
$^{1}$ Center for Computational Sciences, University of Tsukuba, 1-1-1 
 Tennodai, Tsukuba 305-8577 Ibaraki, Japan  
}
\begin{document}

\date{Accepted . Received ; in original form }

\pagerange{\pageref{firstpage}--\pageref{lastpage}} \pubyear{2009}

\maketitle

\label{firstpage}

\begin{abstract}
More than half of nearby disc galaxies have pseudobulges, 
instead of classical bulges that are though to be end-products 
of galaxy mergers. Pseudobulges are presumed to develop over 
time as a result of secular evolution of galaxy discs. 
We report simulations of galaxy formation, in which 
two disc galaxies with disky pseudobulges have formed. 
Based on the profile decomposition, the bulge-to-total mass ratio of 
the simulated galaxies is 0.6 for one galaxy and 0.3 for the other. 
We find that the main formation mechanism of the pseudobulges in our 
simulations is not the secular evolution of discs but high-redshift 
starbursts. 
The progenitors of the pseudobulges form as high-redshift discs with small 
scale lengths by rapid supply of low angular momentum gas. 
By redshift 2, before the main disc formation, pseudobulge 
formation has largely completed in terms of mass. 
The secular evolution such as bar instability accounts for about
30 \% of the bulge mass for one galaxy and only $\sim 13$ \% for 
the other but does affect the final shape and kinematic properties of 
the pseudobulges. 
\end{abstract}


\section{Introduction}

Pseudobulges are characterised by distinctive features from 
classical bulges that are thought to be merger remnants 
\citep{kk04, naab06, hopkins10}. 
Pseudobulges often show substantial rotational support 
and disky or boxy/peanut edge-on shapes. 
These features support the idea that pseudobulges have 
formed via secular evolution of discs \citep{kk04} caused 
by non-asymmetric structure in a galaxy disc such as spiral 
arms and a bar \citep[e.g.][]{combes81, pfenniger90, combes93, 
debattista04, athanassoula05} or by accretion of clumps formed 
in a gas rich unstable disc \citep{noguchi99, is12}. 
We call evolution that is originated by internal processes  
`secular evolution' in this paper regardless of its timescale 
in order to distinguish evolution driven by external processes 
such as mergers\footnote{Indeed, secular evolution can have timescale 
much shorter than a Hubble time \citep[see][for a review]{kk04}. 
\citet{is12} showed that the formation timescale of a clump origin 
pseudobulge is rather short and they distinguished it 
from secular evolution.}. \\ 

In the standard picture of galaxy formation, galaxies form via 
hierarchical merging that naturally produces classical bulges. 
Observationally however more than half of bulges of nearby large 
disc galaxies are pseudobulges \citep{gw08, weinzirl09, kormendy10}. 
This observational fact could be a challenge to the 
standard cosmology \citep{kormendy10, peebles10}. 
Contrarily to the intuition, cosmological simulations of 
Milky Way-sized galaxy formation often find that bulges show surface 
brightness profiles with the S\'ersic indices smaller than 2 
\citep{oka05, gov07, scannapieco11, eris};    
such bulges are usually classified as pseudobulges 
\citep{fisher-drory08, fisher-drory10, weinzirl09}. 
How these bulges form over the cosmic time is however remain 
unclear. 
\\

Investigating formation processes of the simulated bulges of disc 
galaxies should provide important clues to 
understand why pseudobulges are so common both in 
the local Universe and simulated universes. 
Simulations for this purpose must resolve detailed structure of 
galaxies, such as shape of bulges.  
Only recently cosmological simulations with sufficient 
resolution have become possible \citep{of09, ofjt10, eris}. \\

The simulations presented in this paper include a number of 
baryonic processes known to be relevant to galaxy formation, 
such as radiative gas cooling, star formation, supernova feedback, 
and chemical evolution. 
As described in \citet{oka05} and \citet{scannapieco11}, 
there are many uncertainties in modelling of these `sub-grid' 
physics. In particular, feedback is one of the most poorly 
understood  processes,  while it most strongly affects the 
properties of simulated galaxies \citep{oka05, ofjt10, owls}. 
We make use of a model that has 
already had success in reproducing some properties of 
the Local Group satellite galaxies, including the luminosity 
function and luminosity-metallicity relation \citep{ofjt10}
and observed properties of high-redshift Lyman-$\alpha$ 
emitters \citep{shimizu11} and sub-mm galaxies \citep{shimizu12}. \\

The paper is organised as follows. In Section~2, we describe 
our simulations, providing brief descriptions of our modelling 
of baryonic processes. 
In Section~3, we first present 
properties of simulated bulges at redshift 0 to show that 
they have the characteristics of pseudobulges.  
We then show redshift evolution of the simulated galaxies 
and their bulges in order to address formation process of 
the pseudobulges. 
We compare our results with observations in Section~4.  
Finally, in Section~5, we discuss the results and summarise 
our main conclusions. 

\section{The simulations}

To investigate the properties of Milky Way-sized galaxies, 
we select two of the six haloes from the Aquarius project
\citep{aquarius}: `Aq-C' and `Aq-D' in their labelling 
system.  We hence assume a $\Lambda$CDM cosmology with 
the following parameters: $\Omega_{\rm 0} = 0.25, \ 
\Omega_{\Lambda} = 0.75, \ \Omega_{\rm b} = 0.045, \ 
\sigma_8 = 0.9, \ n_{\rm s} = 0.9$, 
and a Hubble constant of $H_0 = 100~h~{\rm km}~{\rm s^{-1}}$~Mpc$^{-1}$, 
where $h = 0.73$. These parameters are consistent with the 
WMAP 1- and 5-year results at the 3~$\sigma$ 
level \citep{wmap1, wmap5}. 
These haloes are extracted from a cosmological periodic 
box of the side length of 100~$h^{-1}$~Mpc, and were chosen to have 
masses close to that of the Milky Way's dark matter halo 
($\sim 10^{12}~M_\odot$). 
These haloes are relatively isolated and do not have 
neighbours exceeding half its mass within $1~h^{-1}$~Mpc 
\citep{navarro10}.\\ 

As in Aquarius, we use the resimulation technique; 
the initial density field of the parent simulation is 
recreated by adding appropriate additional short wavelength 
perturbations in the region out of which the halo of 
interest forms. 
In this region, we also place gas particles which are used 
to perform Smoothed Particle Hydrodynamic (SPH) calculation. 
The region external to this is populated with high-mass dark 
matter particles, the function of which is to reproduce 
the appropriate tidal fields. \\

The high-resolution dark matter particle mass is 
$2.6 \times 10^5~M_\odot$ and $2.2 \times 10^5~M_\odot$, 
  respectively for Aq-C and Aq-D, and  
the original SPH particle mass is thus 
$5.8 \times 10^4~M_\odot$ and $4.8 \times 10^4~M_\odot$. 
Note that the SPH particle mass can change owing to star 
formation and feedback. 
Gravitational softening length is kept constant in comoving 
coordinates as $\epsilon = 1.03$~kpc until redshift 3, and then 
fixed in physical coordinates as $\epsilon = 0.257$~kpc in 
both simulations.  We adopt the same softening length for 
the high-resolution dark matter, gas, and star particles. 

\subsection{The simulation code}

The simulation code is based on an early version of 
a Tree-PM SPH code, {\scriptsize GADGET-3} and it is exactly 
the same as the one used in \citet{parry12}. 
The baryonic processes are modelled so as to reproduce 
the properties of the Local Group satellite 
galaxies \citep{of09, ofjt10}. 
The simulations include metallicity dependent radiative 
cooling and photoheating \citep{wss09} in the presence of 
a time-evolving, spatially uniform ultraviolet background 
\citep[][see also \citealt{ogt08}]{hm01}. 
The star formation occurs in dense gas with $n_{\rm H} > n_{\rm th}$, 
where $n_{\rm th}$ is the threshold gas density above which 
the star formation is enabled. 
The star formation efficiency is set to reproduce the Kennicutt relation 
\citep{ken98} as described in \citet{onb08} \citep[see also G3-TO model 
in][]{aquila}. 
The simulations include both Type II and Ia supernovae 
and follow the evolution of the chemical elements produced 
by these two types of supernovae and by AGB stars. \\

Energetic feedback from supernovae explosion is modelled as 
winds \citep{sh03, od06, onb08, ds08}. 
When a gas particle receives an amount of energy $\Delta E$ 
from Type II supernovae during a timestep, $\Delta t$, 
this particle is added to winds with a probability, 
$p_{\rm w} = \Delta E/(\frac{1}{2}m_{\rm gas} v_{\rm w}^2)$, 
where $m_{\rm gas}$ is the mass of the SPH particle and $v_{\rm w}$ 
is the initial wind speed. 
This is given as $v_{\rm w} = 5 \sigma$, where $\sigma$ is 
the one-dimensional velocity dispersion of the dark matter 
particles around the gas particle \citep[vw5$\sigma$ model in][]{ofjt10}.  
The wind particles are decoupled from the hydrodynamic calculation for 
a short time in order to allow them to escape the high-density star-forming 
regions. When the density has fallen to $n_{\rm H} = 0.01$~ cm$^{-3}$, 
the particles feel the usual hydrodynamic force again. 
If they do not reach sufficiently low density after a time 10 kpc/$v_{\rm w}$, 
they are recoupled to the hydrodynamic interactions anyway.\\

This expression of the wind mass-loading implies that the wind 
mass generated by a Type II supernova is proportional to $\sigma^{-2}$, 
i.e. per unit of star formation, less massive 
galaxies blow more winds than their more massive counterparts. 
This is a key feature to reproduce properties of the Local 
Group satellites, including the shape and approximate 
normalisation of their luminosity function and the 
luminosity-metallicity relation \citep{ofjt10}. \\

The model for supernovae driven winds used here differs 
from the original one by \citet{ofjt10} in two ways. 
Firstly, we allow all gas particles, not just those above 
the star formation threshold density, to be added to the wind 
if they receive feedback energy. 
The original prescription can result in a variable wind mass-loading 
depending on how well the star-forming region is resolved. 
Secondly, only type II supernovae contribute to the winds. 
Type Ia supernova energy is added to the gas as thermal energy. \\

There are several more changes to the code from the one used in 
the previous studies. 
Firstly, the star formation threshold density is now 
a function of the numerical resolution while it was fixed to 
$n_{\rm H} = 0.1$~cc$^{-1}$ in the previous studies.  
For the resolution of the simulations presented in this 
paper, we employ $n_{\rm H} = 1.6$~cc$^{-1}$ as the star formation 
threshold density (see Appendix.~\ref{appendix:c} for more details). 
Secondly, we have implemented a timestep limiter \citep{sm09} 
that reduces the timestep of a gas particle if it is too long 
compared to the neighbouring gas particles.  
Without this limiter, halo gas particles which have much longer 
timesteps than high-speed wind particles often do not feel the force 
from the wind particles (i.e. do not notice the wind particles), 
while the wind particles feel the drag from the surrounding halo 
gas particles. 
Lastly, we have added artificial conductivity in order to capture the 
instabilities at contact surfaces \citep{price08} together with 
the time dependent artificial viscosity \citep{mm97}; otherwise SPH 
cannot deal with such instabilities \citep{oka03, agertz07}.
All these changes have significantly improved the numerical 
convergence as we show in Appendix \ref{appendix:c} \citep[see also][]{parry12}.

\section{Results}

In this section, we first analyse the properties of the simulated 
bulges in order to show that the simulated disc galaxies 
have pseudobulges. 
We then show the evolution of the galaxies and their bulges to 
understand pseudobulge formation.  
The global properties of the simulated galaxies at redshift 0 are summarised in 
Appendix~\ref{appendix:a}.  
In the following analyses, the $z$-direction is chosen to be parallel to the 
angular momentum vector of stars within 5 \% of the virial radius\footnote{The
virial radius is calculated based on the spherical collapse model \citep{ecf96}} 
at given redshift. 
Note that the direction of the angular momentum vector significantly changes
with redshift \citep{oka05, sca09, sales12} as we will show later. 
The $x$-direction is to be parallel to the major axis of the surface stellar density 
distribution in face-on. 
We also define galaxy radius as 10 \% of the virial radius. 

\subsection{Shapes and kinematic properties of the bulges}

\begin{figure}
  \begin{center}
    \includegraphics[width=8.0cm]{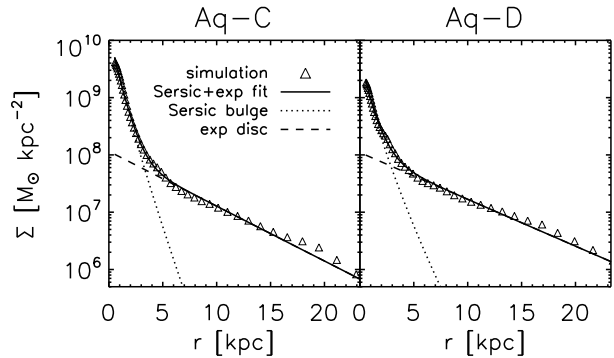}
  \end{center}
  \caption{Azimuthally averaged surface stellar density profiles. 
    The left and right panels show profiles of Aq-C and Aq-D galaxies, respectively.  
    The profiles of each galaxy is fitted by a combination of the S\'ersic and 
    exponential profiles. 
    The surface density profiles are indicated by the triangles and the best fit 
    profiles are shown by the solid lines. 
    The contribution from S\'ersic bulges (dotted lines) and the exponential discs 
    (dashed lines) are also plotted. 
    The bulges are well fitted by the S\'ersic indices, $n \simeq 1.2$ and $1.4$, for 
    Aq-C and Aq-D, respectively. 
  }
  \label{fig:profiles}
\end{figure}

For classifying bulges, the S\'ersic profile fitting for a surface stellar 
density profile is frequently used:
\begin{equation}
\Sigma(r) = \Sigma_e \exp\left[
  -b_n \left\{\left(\frac{r}{R_e}\right)^\frac{1}{n} - 1.0 \right\}\right
  ], 
  \label{eq:sersic}
\end{equation}
where $R_e$ is the effective radius and $\Sigma_{\rm e}$ 
is the surface density at $R_{\rm e}$, respectively, 
and $n$ is the S\'ersic index. 
The parameter $b_n$ is related to $n$ by $b_n \simeq 2 n - 0.324$.  
Bulges with $n < 2$ are usually classified as 
pseudobulges 
\citep{kk04, fisher-drory08, fisher-drory10, weinzirl09}\footnote{Note 
that this classification at $n = 2$ was questioned by \citet{graham11} 
because bulges ranging from $n = 0.5$ to 10 obey a single curved 
relation in the $R_e-n$ plane.}.
In Fig.~\ref{fig:profiles}, we show the surface density profiles 
of the simulated galaxies from face-on ($x$--$y$ projection) 
at redshift 0. 
We fit each profile by a combination of the S\'ersic bulge and 
exponential disc profiles. 
We obtain $n \simeq 1.2$ and $1.4$ for the bulges of Aq-C and Aq-D, 
respectively. 
These bulges hence have pseudobulge-like profiles. 
According to the profile fitting, 
the bulge-to-total mass ratios, $B/T$, are $0.6$ for Aq-C and $0.3$ 
for Aq-D. 
These values suggest that Aq-C is an early-type disc galaxy and 
Aq-D is presumably classified as an Sb galaxy.  
The bulge stars dominate mass 
within 3~kpc from the galaxy centres; more than 92 \% and 
86 \% for Aq-C and Aq-D, respectively.
We thus regard this region as a bulge. \\

\begin{figure}
  \begin{center}
    \includegraphics[width=8.0cm]{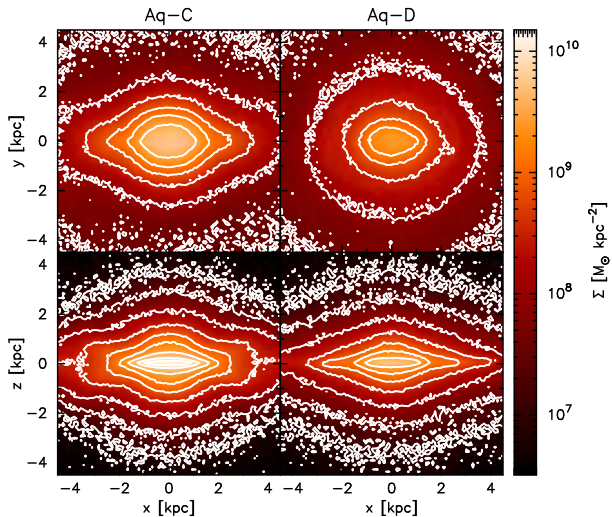}
  \end{center}
  \caption{
    Surface stellar density maps of the bulges. 
    The upper panels show the face-on views ($x$--$y$ projection) 
    and the lower panels show the edge-on views ($x$--$z$ projection).  
    The contour levels are chosen to highlight the bulge shapes, 
    which are clearly `disky' in the edge-on views. 
  }
  \label{fig:denmap}
\end{figure}

In Fig.\ref{fig:denmap}, we show the face-on (upper panels) and 
edge-on (lower panels) surface stellar density maps.  
From the face-on density maps, it is evident that Aq-C has a bar, while  
Aq-D shows almost axisymmetric surface stellar density distribution except in 
the central region ($r < 2$~kpc) where we see weak non-axisymmetricity.  
In the edge-on density maps, both bulges  
appear to be quite `flat'. 
This is not what one would expect for a classical bulges;  
both galaxies seem to posses `disc-like' bulges 
\citep[][or `disky pseudobulges' in \citealt{erwin08}'s terminology]{athanassoula05}. 
A further evidence for a dominant central disk-like structure is the 
diamond shaped isodensity contours in Aq-D's bulge \citep{pohlen00}.  
The inner part of Aq-C's bulge also shows disky, diamond isodensity contours, 
while the outer part has `boxy' features ($2 \lesssim |x| \lesssim 4.5$~kpc 
and $1 \lesssim |z| \lesssim 2$~kpc). 
The boxy/peanut shape is related to bar structure \citep{athanassoula05} 
and therefore the boxy isodensity contours are consistent with 
the fact that Aq-C has a bar. \\

\begin{figure}
    \includegraphics[width=8.0cm]{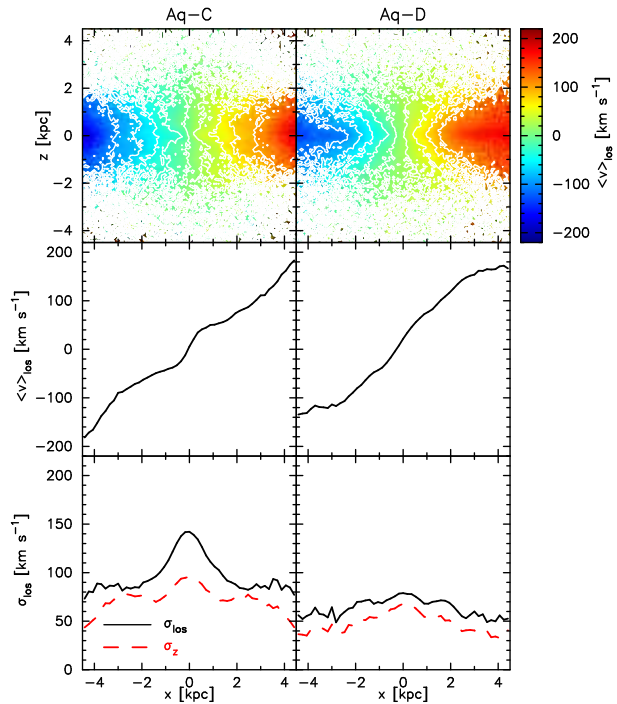}
  \caption{
    Kinematic properties of the bulges from edge-on view ($x$--$z$ projection). 
    {\it Top panels}: The mean line-of-sight velocity maps. 
    The contour levels correspond to 
    $|\langle v_{\rm los} \rangle| = 0, \ 25, \ 50, 
    \ 75, \ 100, \ 150 \ {\rm km}\ {\rm s}^{-1}$. 
    {\it Middle panels}: The line-of-site velocity profiles at the disc 
    plane ($|z| < 0.1$~kpc). 
    {\it Bottom panels}: The line-of-sight velocity dispersion profiles at 
    the disc plane. 
    We also show the vertical ($z$-direction) velocity dispersion profiles at 
    the disc plane by the red dashed lines. 
  }
  \label{fig:velmap}
\end{figure}
Next we investigate the kinematic properties of the simulated bulges.
In the top panels of Fig.~\ref{fig:velmap}, we show the mean line-of-sight 
velocity maps from edge-on view. 
Both bulges have significant rotation but the shapes of the isovelocity 
contours are rather different between them; Aq-C's contours are more vertical 
than Aq-D's, implying the rotation properties of Aq-C's bulge is more cylindrical. 
This is presumably due to the presence of the bar. \\

We plot the line-of-sight velocity profiles (rotation curves) at the 
disc plane ($|z| < 0.1$~kpc) in the middle panels and the line-of-sight 
velocity dispersion profiles in the bottom panels. 
We find that the inner part of Aq-C's bulge is rather kinematically hot; 
the velocity dispersion profiles are sharply peaked at $x = 0$. 
The region is supported by the velocity dispersion of the stars rather 
than the rotation. 
We also plot the vertical ($z$-direction) velocity dispersion of the 
stars in the disc plane. The vertical velocity dispersion is only half 
the line-of-sight one. 
Therefore the orbits of stars in the inner part of the bulge 
are largely non-circular but confined in a thin oblate. 
On the other hand, Aq-D's bulge is kinematically cold. 
The line-of-sight velocity dispersion profile is very flat 
and it seems typical of the disky pseudobulges \citep{erwin08}. 
The cold, highly rotationally supported feature of the bulge 
implies that the  pseudobulge is a disc-like bulge. 

\subsection{Evolution of the galaxies and their bulges}

\begin{figure*}
  \begin{center}
    \includegraphics[width=16.0cm]{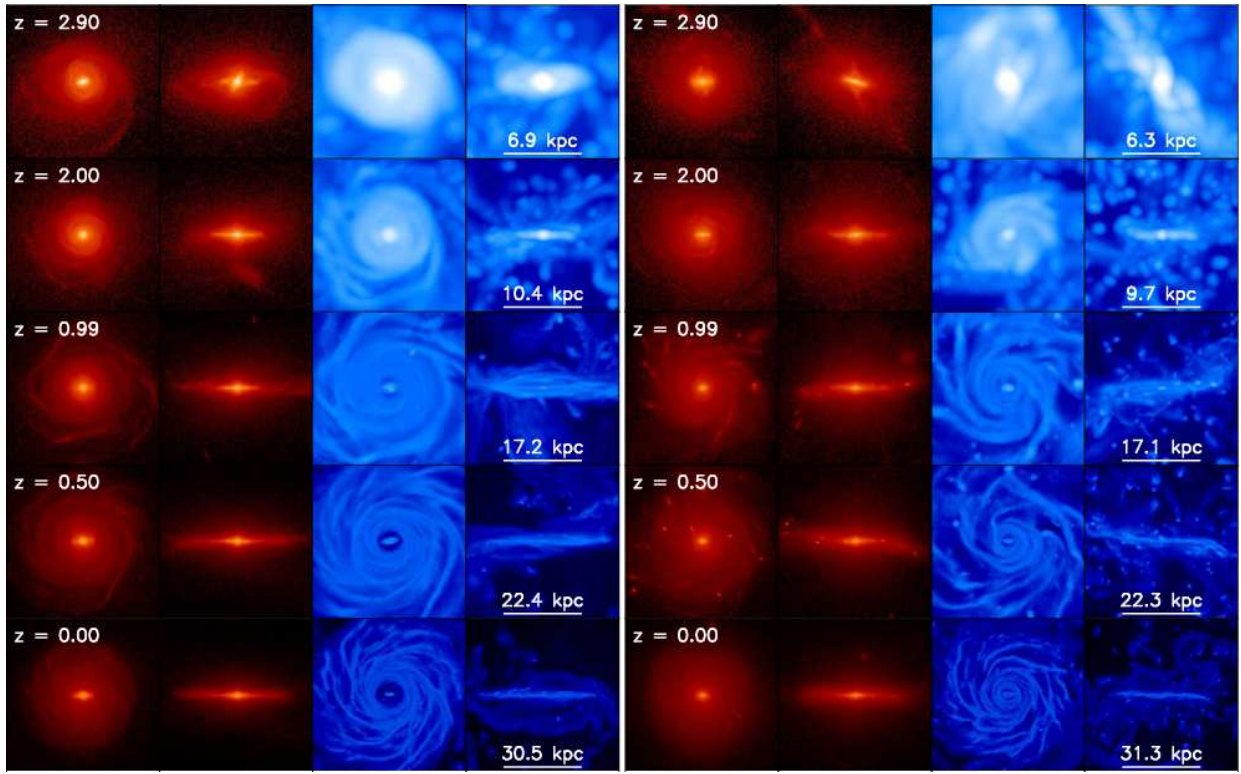}
  \end{center}
  \caption{Evolution of the galaxies. 
    The side length of each panel is set to 20 \% of the virial 
    radius of the host dark matter halo at a given redshift.  
    The galaxy radius which is defined as 10 \% of the virial 
    radius is indicated by a horizontal line in each panel. 
    The left and right concatenated panels, respectively, show Aq-C and Aq-D 
    galaxies. From left to right, the surface stellar density in 
    face-on ($x$-$y$ projection) and in edge-on ($x$-$z$), 
    and surface gas density in face-on and in edge-on are 
    presented. 
    Brighter colour is used for higher density. 
    From top to bottom, we show the main progenitor at redshift 
    2.9, 2.0, 0.99, 0.5, and 0. 
    The $z$-direction is chosen to be parallel to the angular 
    momentum of stars within 5 \% of the virial radius at 
    each redshift. 
    The $x$-direction is to be parallel to the major axis 
    as in Fig.~\ref{fig:denmap} and \ref{fig:velmap}. 
  }
  \label{fig:snaps}
\end{figure*}

In this subsection we investigate how the galaxies and their 
bulges form and evolve in the simulations. 
Fig.~\ref{fig:snaps} shows stellar and gas distribution 
in the main progenitors of Aq-C and Aq-D galaxies around 
redshift 3, 2, 1, 0.5, and 0. Again we chose the 
$z$-direction to be parallel to the angular momentum 
vector and $x$-direction to the major axis. 
At redshift 3, both galaxies show irregular morphology, 
characterised by an inner disc and misaligned outer 
disc or ring. The misalignment is caused 
by the change of the direction of angular momentum vector 
of newly accreted gas as we will show later. 
The components that eventually evolve into the discs at 
redshift 0 start forming around redshift 2. 
Aq-C forms a bar around redshift 1 and the bar survives 
until redshift 0. 
In Aq-D, a weak (and short) bar-like feature can be seen 
at redshift 2, which disappears by redshift 1. 
Aq-D's late-time evolution below redshift 1 is characterised 
by clumpy star formation. \\

\begin{figure}
  \begin{center}
    \includegraphics[width=8.0cm]{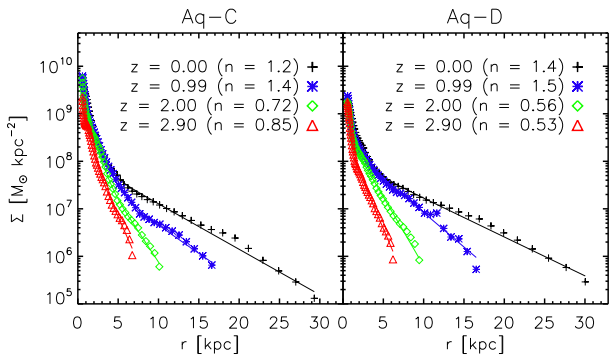}
  \end{center}
  \caption{Evolution of the surface stellar density profiles.  
    The surface stellar density profiles of Aq-C (left) and Aq-D (right) 
    galaxies and their main progenitors obtained from face-on views 
    are presented. 
    The profiles around redshift 0, 1, 2, and 3 are indicated by the black 
    plus signs, blue asterisks, green diamonds, and red triangles, 
    respectively. 
    Each profile is fitted by a combination of a S\'ersic bulge and 
    an exponential disk, which is shown by the line. 
    The S\'ersic indices are shown in each panel. 
  }
  \label{fig:surfevo}
\end{figure}

Our main result is depicted in Fig.~\ref{fig:surfevo} where  
we show the surface stellar density profiles of the main 
progenitor galaxies from face-on view at several redshifts.  
Redshift evolution of the surface density profiles indicates that 
a high-redshift disc with a small scale length forms first, 
then a component that eventually evolves into the disc at redshift 0 
develops around it. 
In fact, the S\'ersic bulges are always disc-like ($n < 2$).   
It is evident that there has been only a little evolution in the bulge mass 
since redshift 2; the bulge mass at redshift 2 already accounts for 
$\sim 70$ \% of that at redshift 0 for Aq-C and $\sim 87$ \% for Aq-D. 
Because the discs seen at redshift 0 start forming around redshift 2, 
we conclude that the main channel of the pseudobulge formation in 
our simulations is not secular evolution of the main (or outer) discs, 
although secular evolution seems to have non-negligible contribution 
($\sim 30$ \%) to the mass of Aq-C's bulge, which is the one 
that has the bar at low redshift.  
\\

\begin{figure}
  \begin{center}
    \includegraphics[width=8.0cm]{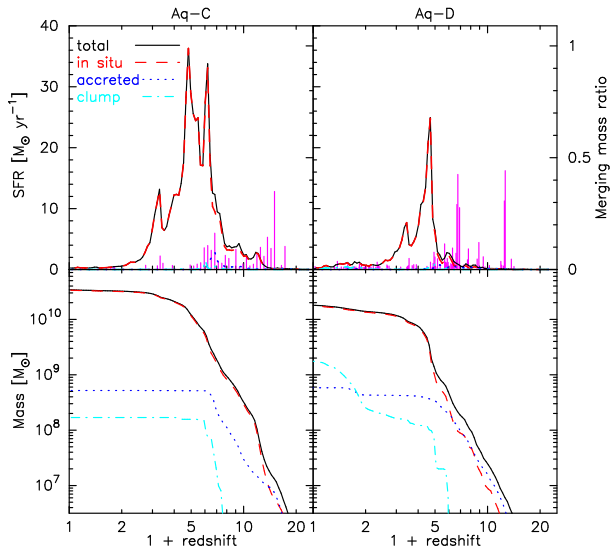}
  \end{center}
  \caption{Distribution of formation times of bulge stars and merging histories 
    of galaxies expressed in terms of redshift. 
    Stars lie within 3~kpc from the galaxy centre at 
    redshift 0 are identified as the bulge stars here. 
    The left and right panels show Aq-C and Aq-D galaxies, respectively. 
    The upper and lower panels show the same data in differential and 
    cumulative form, respectively.
    The black solid lines indicate all the bulge stars, while 
    the red dashed and blue dotted lines respectively indicate those 
    born in the main progenitors ({\it in situ}) and those born 
    in the satellites ({\it accreted}). 
    The cyan dot-dashed lines show in situ stars which have formed in 
    gas clumps.
    In the upper panels, we show the merging mass ratios by the vertical 
    magenta lines.  
    We show all the mergers whose mass ratios are greater than $0.01$. 
  } 
  \label{fig:bulgeform}
\end{figure}

In Fig.~\ref{fig:bulgeform}, we present the formation histories of bulge 
stars:  the stars lie within 3~kpc from the galaxy centres at redshift 0. 
In order to distinguish the stars that are born in the main progenitor
({\it in situ}) from the stars that are brought by accreted satellites 
({\it accreted}), 
we identify subhaloes by using {\scriptsize SUBFIND} \citep{spr01} and 
construct merger trees. 
We define subhaloes as systems that consists of at least 32 
self-bound particles. 
To identify clumps formed by the self-gravitational instability of gas 
(see Aq-D in Fig.~\ref{fig:snaps}), 
we also utilise {\scriptsize SUBFIND}; 
a group of self-bound particles that contains more than 32 particles 
found within the galaxy radius, $0.1 r_{\rm vir}$, whose baryon 
fraction is more than 90 \% is regarded as a clump. 
\\

As shown in the top panels of Fig.~\ref{fig:bulgeform}, 
most of the bulge stars are born in starbursts between 
redshifts 2 and 6 in Aq-C and between 2 and 4 in Aq-D. 
Moreover, most of the bulge stars are in situ. 
This fact implies that at least `dry mergers' do not contribute to 
the formation of the bulges. 
There is little star formation activity below redshift 1 in Aq-C's 
bulge, while Aq-D's bulge show slight activity at low redshift. 
We find that about half the star formation in  Aq-D's bulge  
below redshift 1 occurs in the gas clumps.
The clumps sink to the galaxy centre by dynamical friction and 
add mass to the bulge \citep{is12}.  
However, the contribution from the clumps is only 
about 10 \% and not as large as in 
the `clump origin pseudobulge' reported by \citet{is12}. 
\\

In order to investigate what triggers the starbursts, we study 
the merging histories of the galaxies. 
We define galaxy mass, $M_{\rm gal}$, as the sum of the 
stellar mass, $M_*$, and the interstellar medium (ISM) mass,  
$M_{\rm ism}$, in a subhalo, were we regard the gas with high 
density ($n_{\rm H} > 0.1~{\rm cc}^{-1}$) and low temperature 
($T < 3\times10^4$~K) as the ISM. The star-forming gas 
($n_{\rm H}  > 1.6~{\rm cc}^{-1}$) is also counted as the ISM.  
The merging mass ratio is defined by the second massive progenitor 
to the main progenitor mass in a merger (the vertical lines in 
the upper panels of Fig.~\ref{fig:bulgeform}). 
\\

We find that the star formation activity is not strongly correlated 
with mergers. We thus conclude that the starbursts are triggered by the 
rapid supply of gas at high redshift instead of the mergers 
(see Appendix~\ref{appendix:b}). 
We also find that both galaxies have not undergone mergers with mass
ratio greater than 0.1 since redshift 4. 
We have confirmed that the major mergers at high redshift are {\it wet} 
(Appendix~\ref{appendix:b}); 
wet mergers are not as destructive as dry ones \citep{hopkins10} 
and sometimes form discs by depositing orbital angular momentum 
\citep{sh05, robertson06, gov09}. 
The gas richness at high redshift seems to prevent the galaxies from 
forming classical bulges. 
\\

\begin{figure}
  \begin{center}
    \includegraphics[width=8.0cm]{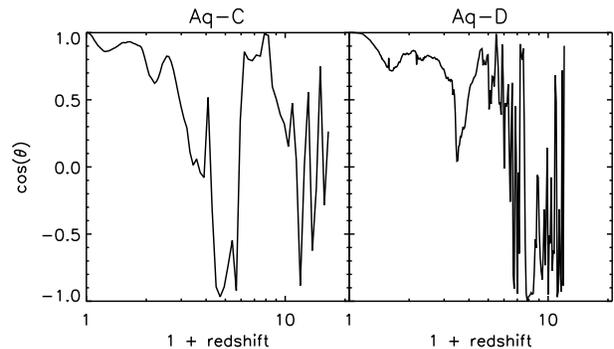}
  \end{center}
  \caption{Spin flips: Evolution of the orientation of the galaxies. 
    The cosine of the angle between the direction of the spin 
    of the galaxy at redshift 0 and that of the main progenitor at each 
    redshift is plotted as a function of redshift in each panel. 
    The direction is computed using the stellar component 
    within $0.5 r_{\rm vir}$, where stars in the satellites 
    are removed from the analysis. 
  }
  \label{fig:angleevo}
\end{figure}

Finally, we study evolution of the orientation of the main progenitor 
galaxies. 
During the growth of a dark halo, the direction of its spin undergoes 
rapid changes \citep{bf12}. The direction of the angular momentum vector 
of newly accreting gas also changes dramatically. This induces the `flip' of 
a disc \citep{oka05} and sometimes destroyed a preexisting disc 
\citep{scannapieco11}. 
\citet{sales12} pointed out, by using {\scriptsize GIMIC} 
simulations \citep{gimic}, that many bulges in Milky Way-sized haloes form 
owing to this misalignment of newly-accreted gas. \\

In Fig.\ref{fig:angleevo}, we show the evolution of the galaxy orientation. 
We define the spin of a disc by using the direction of the angular momentum 
vector of the stellar component within $0.5 r_{\rm vir}$, where we ignore 
the stars in satellites to avoid misdefining the galaxy orientation due 
to the orbital angular momentum of the satellites. 
The orientation of the galaxies rapidly changes at high redshift by the 
change of the direction of the angular momentum vector of newly-accreted  
materials. 
\\

The flips displayed in Fig.~\ref{fig:angleevo} cause to form the misaligned discs 
for example seen at redshift 3 in Fig.~\ref{fig:snaps} 
{
  The rapid change of the spin direction at high redshift ($ > 2$) prevents 
  disc formation and forms pseudobulges.  
  At redshift below 2, at which the host haloes are well established, 
  the flip is subsided and the discs develop around the pseudobulges. 
}
\\

\section{Comparison with observations} \label{sec:obs}
  In this section, we compare the simulated galaxies and bulges 
  with observations. We firstly investigate whether the simulated 
  galaxies are too bulge-dominated to host pseudobulges or not.  
  Galaxies with pseudobulges are in general 
  have smaller $B/T$ than those with classical bulges 
  \citep[e.g.][]{gw08, fisher-drory08, fisher+09, weinzirl09}. 
  For example \citet{fisher-drory08} showed that most of the 
  galaxies with pseudobulges have $B/T < 0.4$ in $V$-band and 
  $\simeq 0.5$ at a maximum; \citet{fisher+09} reported that 
  such galaxies have $B/T < 0.6$ in near-IR;  
  All high-mass ($M_* \ge 10^{10} M_\odot$) spirals whose bulge 
  indices are smaller than 2 have $H$-band $B/T \lesssim 0.4$ 
  \citep{weinzirl09}. 
  In our simulations, the values of $H$-band $B/T$ are {0.46 and 0.21} 
  for Aq-C and Aq-D, respectively\footnote{$B/T$ becomes smaller in bluer 
  bands (see also Table.~\ref{table:gal})}. 
  Therefore Aq-C's $B/T$ value is on the very high end of what has been 
  observed and it might be out of range because we ignore the dust extinction 
  which reduces the disc luminosity. \\

  Next, we compare the specific star formation rate in the bulges with 
  observed values. 
  Following \citet{fisher+09}, the bulge radius, $R_{\rm XS}$, is defined 
  as the radius at which the surface stellar density profile of a bulge 
  is 25 \% higher than that of a disc. The bulge mass, $M_{\rm XS}$, is 
  then given as:
\begin{equation} 
  M_{\rm XS} = M_{\rm star}(<R_{\rm XS}) 
  - 2 \pi \Sigma_0 \int_0^{R_{\rm XS}} r e^{-r/R_{\rm d}} \dm r.   
\end{equation}
  We list the bulge mass, the bulge radius, the star formation rate internal 
  to the bulge, $\dot{M}_{\rm XS}$, and the specific star formation rate of the bulge, 
  $\dot{M}_{\rm XS}/M_{\rm XS}$, in Table~\ref{table:ssfr}. 
  The specific star formation rates, $1.1 \times 10^{-11}$~yr$^{-1}$ (Aq-C) 
  and $1.4 \times 10^{-11}$~yr$^{-1}$ (Aq-D), are in good agreement with 
  those for the {\it inactive} pseudobulges \citep{fisher+09}. 
  Pseudobulge formation via high-redshift starbursts provides a good explanation 
  for the formation of inactive pseudobulges. \\

  Bulges of nearby galaxies are however less massive than $10^{10}~M_\odot$ 
  \citep{fisher+09}.
  Aq-C's bulge ($M_{\rm XS} = 2.2 \times 10^{10}$) seems to be too massive.  
  This is because that the simulated galaxies have too much star formation 
  at high redshift. 
  Aq-C has formed half of the final stellar mass by redshift 3.3 and 
  Aq-D by redshift 1.8. 
  \citet{leitner12} argued that current star-forming galaxies formed most of 
  their stellar mass at redshift lower than 2.  
  Results from abundance matching also show that Aq-C forms too many stars 
  at redshift higher than 2 \citep[][their Fig.~12]{moster12} for its halo mass.  
  The disagreement at high redshift indicates possible failure of our galaxy 
  formation model. We will discuss effects of this early star formation in 
  Section~\ref{sec:discussion}.  
\begin{table}
\caption{Size, mass  and star formation rate of the simulated bulges at redshift 0.}
\label{table:ssfr}
\begin{tabular}{@{}lcccc}
\hline
Galaxy 
  & $R_{\rm XS}$ 
  & $M_{\rm XS}$ 
  & $\dot{M}_{\rm XS}$
  & $\dot{M}_{\rm XS}/M_{\rm XS}$ \\
  & (kpc) & ($M_\odot$) & ($M_\odot~{\rm yr}^{-1}$) & (yr$^{-1}$) \\
\hline
Aq-C & $3.2$ & $2.2 \times 10^{10}$ & $2.4 \times 10^{-1}$ & $1.1 \times 10^{-11}$ \\
Aq-D & $2.7$ & $7.6 \times 10^9$    & $1.1 \times 10^{-1}$ & $1.4 \times 10^{-11}$ \\
\hline
\end{tabular}
\end{table}

\section{Summary and discussion} \label{sec:discussion}

We have performed $N$-body/hydrodynamic simulations of galaxy formation 
in two Milky Way-sized haloes in a $\Lambda$CDM universe. 
The haloes are taken from the Aquarius project \citep{aquarius}: Aq-C and Aq-D. 
Both galaxies form well defined discs; one has the bulge-to-total mass ratio, 
$B/T \simeq 0.6$, and the other, $B/T \simeq 0.3$, based on the surface 
stellar density profiles.  
The S\'ersic index of Aq-C's bulge is 1.2 and that of Aq-D's bulge is 
1.4. These values suggest that both bulges are `pseudobulges'. 
The edge-on surface stellar density maps confirm that these bulges 
indeed show the characteristics of pseudobulges, they particularly  
resemble `disky pseudobulges', while Aq-C also has features of a `boxy bulge'.  
The boxiness is consistent with the existence of a bar in Aq-C 
\citep{athanassoula05}. \\

The formation histories of these pseudobulges are rather different from the 
standard picture, in which pseudobulges are believed to form through secular 
evolution of galaxy discs. 
In our simulations, the pseudobulges mainly form by high-redshift starbursts 
before redshift 2.  
The evolution of the surface stellar density profiles reveals that the pseudobulge 
components are already in place at redshift 2--3 as disky components with 
small scale length. The mass of these central disky components at redshift 2 
accounts for $\sim 70$\% and $\sim 87$\% of the final pseudobulge mass of Aq-C and Aq-D, 
respectively. 
These progenitors of the pseudobulges would be observed as high-redshift discs  
\citep{stark08}.
In fact, when the galaxies have peak star formation rates, the star formation occurs 
in unperturbed gas discs. 
The high-redshift discs continuously change the orientation by exchanging the angular 
momentum with newly-accreted materials whose angular momentum is misaligned with 
the preexisting discs. 
In the course of the frequent flips, the discs are heated and turn into 
pseudobulge-like objects.\\

The main stage of the formation of the pseudobulges ends soon after the starbursts 
by gas consumption due to the starbursts and the energetic winds that removes 
star-forming gas. 
When the host haloes are well established, the gas cooled in the haloes has larger 
angular momentum and gradually develops the discs with larger scale length 
than the pseudobulges. 
The secular evolution of the main discs that is mainly driven by bar instability 
in Aq-C and clumpy star formation in Aq-D further increases the mass of the pseudobulges. 
The secular evolution however accounts for only $\sim 30$\% and $\sim 13$\% of the 
mass of the pseudobulges of Aq-C and Aq-D, respectively. 
We therefore conclude that the main channel of the pseudobulge formation in our simulations 
is {\it not} the secular evolution of the discs. \\

  The specific star formation rates of the simulated bulges 
  are consistent with those of the observed {\it inactive} pseudobulge 
\citep{fisher+09}, implying that such pseudobulges have formed via 
high-redshift starbursts. The bulge mass, Aq-C's bulge in particular,  
is however too large to compare with the observed counterparts. 
Aq-C indeed shows too much early star formation before redshift 2 
\citep{moster12}. 
We are thus likely to overestimate the importance of the high-redshift starbursts 
for pseudobulge formation especially for Aq-C.
\\

There seems to be two reasons why the simulated galaxies do not 
form classical bulges. One is that the host haloes have relatively quiet merger 
histories as described in \citet{wang11}. Consequently, the galaxies have not 
undergone mergers with mass ratios greater than 0.1 since redshift 4.
The merging mass ratios of the galaxies becomes smaller than those of their 
host haloes owing to the feedback which is more effective in smaller galaxies.  
The other reason is that when the galaxies are quite gas rich when they undergo 
major mergers at high redshift. This is again due to our feedback that suppresses 
star formation very efficiently when the main progenitors are small. 
Wet (gas rich) mergers are not as destructive as dry (collisionless) mergers 
\citep{hopkins10} and often form discs as merger remnants 
\citep{sh05, robertson06, gov09}. 
Any successful simulations of disc galaxy formation must suppress early 
star formation \citep{brook04, oka05, eris, aquila}; we thus speculate that gas 
richness in high-redshift progenitors is the key to explain the existence 
of inactive pseudobulges in large disc galaxies. 
\\

The formation scenario of pseudobulges by high-redshift starbursts 
presented in this paper provides a good explanation for the existence of 
inactive pseudobulges in early-type disc galaxies such as S0 and 
Sa galaxies. 
Pseudobulges do exist in early-type disc galaxies \citep{kk04} and 
secular evolution may take too long to form such large pseudobulges. 
In our simulations, large inactive pseudobulges with the bulge-to-total 
mass 
ratios, 
$B/T \sim 0.6$ and $\sim 0.3$; $\sim 70$~\% and $\sim 87$~\% of 
them have formed through high-redshift starbursts, respectively, and secular 
evolution adds the rest. 
A simulation of a smaller galaxies $M_{\rm gal} \sim 10^{10}~M_\odot$, 
for which feedback is more efficient than in Milky Way-sized galaxies, 
showed that its pseudobulge has formed by the secular evolution \cite{brook12}. 
We hence expect that different formation processes of pseudobulge operate 
depending on galaxy mass.

\section*{Acknowledgements}
I am grateful to the anonymous referee for helpful comments. 
I would like to thank Shigeki Inoue, Junich Baba, and Daisuke Kawata
for stimulating discussion. 
The simulations were performed with T2K Tsukuba at Centre 
for Computational Sciences in University of Tsukuba.  
This work was supported in part by Grant-in-Aid for Scientific Research 
(S) by JSPS (20224002),  by Grant-in-Aid for Young Scientists 
(B: 24740112), and by MEXT HPCI STRATEGIC PROGRAM. 


\begin{thebibliography}{71}
\expandafter\ifx\csname natexlab\endcsname\relax\def\natexlab#1{#1}\fi

\bibitem[{{Abadi} {et~al.}(2003){Abadi}, {Navarro}, {Steinmetz} \&
  {Eke}}]{aba03b}
{Abadi} M.~G., {Navarro} J.~F., {Steinmetz} M., {Eke} V.~R., 2003, ApJ, 597, 21

\bibitem[{{Agertz} {et~al.}(2007){Agertz}, {Moore}, {Stadel}, {Potter},
  {Miniati}, {Read}, {Mayer}, {Gawryszczak}, {Kravtsov}, {Nordlund}, {Pearce},
  {Quilis}, {Rudd}, {Springel}, {Stone}, {Tasker}, {Teyssier}, {Wadsley} \&
  {Walder}}]{agertz07}
{Agertz} O., {et~al.}, 2007, \mnras, 380, 963

\bibitem[{{Athanassoula}(2005)}]{athanassoula05}
{Athanassoula} E., 2005, \mnras, 358, 1477

\bibitem[{{Behroozi} {et~al.}(2010){Behroozi}, {Conroy} \&
  {Wechsler}}]{behroozi10}
{Behroozi} P.~S., {Conroy} C., {Wechsler} R.~H., 2010, \apj, 717, 379

\bibitem[{{Bett} \& {Frenk}(2012)}]{bf12}
{Bett} P.~E., {Frenk} C.~S., 2012, \mnras, 420, 3324

\bibitem[{{Brook} {et~al.}(2011){Brook}, {Governato}, {Ro{\v s}kar}, {Stinson},
  {Brooks}, {Wadsley}, {Quinn}, {Gibson}, {Snaith}, {Pilkington}, {House} \&
  {Pontzen}}]{brook11}
{Brook} C.~B., {et~al.}, 2011, \mnras, 415, 1051

\bibitem[{{Brook} {et~al.}(2004){Brook}, {Kawata}, {Gibson} \&
  {Flynn}}]{brook04}
{Brook} C.~B., {Kawata} D., {Gibson} B.~K., {Flynn} C., 2004, \mnras, 349, 52

\bibitem[{{Brook} {et~al.}(2012){Brook}, {Stinson}, {Gibson}, {Ro{\v s}kar},
  {Wadsley} \& {Quinn}}]{brook12}
{Brook} C.~B., {Stinson} G., {Gibson} B.~K., {Ro{\v s}kar} R., {Wadsley} J.,
  {Quinn} T., 2012, \mnras, 419, 771

\bibitem[{{Combes} \& {Elmegreen}(1993)}]{combes93}
{Combes} F., {Elmegreen} B.~G., 1993, \aap, 271, 391

\bibitem[{{Combes} \& {Sanders}(1981)}]{combes81}
{Combes} F., {Sanders} R.~H., 1981, \aap, 96, 164

\bibitem[{{Crain} {et~al.}(2009){Crain}, {Theuns}, {Dalla Vecchia}, {Eke},
  {Frenk}, {Jenkins}, {Kay}, {Peacock}, {Pearce}, {Schaye}, {Springel},
  {Thomas}, {White} \& {Wiersma}}]{gimic}
{Crain} R.~A., {et~al.}, 2009, \mnras, 399, 1773

\bibitem[{{Dalla Vecchia} \& {Schaye}(2008)}]{ds08}
{Dalla Vecchia} C., {Schaye} J., 2008, \mnras, 387, 1431

\bibitem[{{Debattista} {et~al.}(2004){Debattista}, {Carollo}, {Mayer} \&
  {Moore}}]{debattista04}
{Debattista} V.~P., {Carollo} C.~M., {Mayer} L., {Moore} B., 2004, \apjl, 604,
  L93

\bibitem[{{Eke} {et~al.}(1996){Eke}, {Cole} \& {Frenk}}]{ecf96}
{Eke} V.~R., {Cole} S., {Frenk} C.~S., 1996, \mnras, 282, 263

\bibitem[{{Erwin}(2008)}]{erwin08}
{Erwin} P., 2008, in IAU Symposium, Vol. 245, IAU Symposium, {M.~Bureau,
  E.~Athanassoula, \& B.~Barbuy}, ed., pp. 113--116

\bibitem[{{Fisher} \& {Drory}(2008)}]{fisher-drory08}
{Fisher} D.~B., {Drory} N., 2008, \aj, 136, 773

\bibitem[{{Fisher} \& {Drory}(2010)}]{fisher-drory10}
---, 2010, \apj, 716, 942

\bibitem[{{Fisher} {et~al.}(2009){Fisher}, {Drory} \& {Fabricius}}]{fisher+09}
{Fisher} D.~B., {Drory} N., {Fabricius} M.~H., 2009, \apj, 697, 630

\bibitem[{{Governato} {et~al.}(2010){Governato}, {Brook}, {Mayer}, {Brooks},
  {Rhee}, {Wadsley}, {Jonsson}, {Willman}, {Stinson}, {Quinn} \&
  {Madau}}]{gov10}
{Governato} F., {et~al.}, 2010, \nat, 463, 203

\bibitem[{{Governato} {et~al.}(2009){Governato}, {Brook}, {Brooks}, {Mayer},
  {Willman}, {Jonsson}, {Stilp}, {Pope}, {Christensen}, {Wadsley} \&
  {Quinn}}]{gov09}
---, 2009, \mnras, 398, 312

\bibitem[{{Governato} {et~al.}(2007){Governato}, {Willman}, {Mayer}, {Brooks},
  {Stinson}, {Valenzuela}, {Wadsley} \& {Quinn}}]{gov07}
{Governato} F., {Willman} B., {Mayer} L., {Brooks} A., {Stinson} G.,
  {Valenzuela} O., {Wadsley} J., {Quinn} T., 2007, \mnras, 374, 1479

\bibitem[{{Graham}(2011)}]{graham11}
{Graham} A.~W., 2011, ArXiv e-prints, 1108.0997

\bibitem[{{Graham} \& {Worley}(2008)}]{gw08}
{Graham} A.~W., {Worley} C.~C., 2008, \mnras, 388, 1708

\bibitem[{{Guedes} {et~al.}(2011){Guedes}, {Callegari}, {Madau} \&
  {Mayer}}]{eris}
{Guedes} J., {Callegari} S., {Madau} P., {Mayer} L., 2011, \apj, 742, 76

\bibitem[{{Guo} {et~al.}(2010){Guo}, {White}, {Li} \& {Boylan-Kolchin}}]{guo10}
{Guo} Q., {White} S., {Li} C., {Boylan-Kolchin} M., 2010, \mnras, 404, 1111

\bibitem[{{Haardt} \& {Madau}(2001)}]{hm01}
{Haardt} F., {Madau} P., 2001, in Clusters of Galaxies and the High Redshift
  Universe Observed in X-rays, {Neumann} D.~M., {Tran} J.~T.~V., eds.

\bibitem[{{Hopkins} {et~al.}(2010){Hopkins}, {Bundy}, {Croton}, {Hernquist},
  {Keres}, {Khochfar}, {Stewart}, {Wetzel} \& {Younger}}]{hopkins10}
{Hopkins} P.~F., {et~al.}, 2010, \apj, 715, 202

\bibitem[{{Inoue} \& {Saitoh}(2012)}]{is12}
{Inoue} S., {Saitoh} T.~R., 2012, \mnras, 422, 1902

\bibitem[{{Kennicutt}(1998)}]{ken98}
{Kennicutt} Jr. R.~C., 1998, ApJ, 498, 541

\bibitem[{{Komatsu} {et~al.}(2009){Komatsu}, {Dunkley}, {Nolta}, {Bennett},
  {Gold}, {Hinshaw}, {Jarosik}, {Larson}, {Limon}, {Page}, {Spergel},
  {Halpern}, {Hill}, {Kogut}, {Meyer}, {Tucker}, {Weiland}, {Wollack} \&
  {Wright}}]{wmap5}
{Komatsu} E., {et~al.}, 2009, \apjs, 180, 330

\bibitem[{{Kormendy} {et~al.}(2010){Kormendy}, {Drory}, {Bender} \&
  {Cornell}}]{kormendy10}
{Kormendy} J., {Drory} N., {Bender} R., {Cornell} M.~E., 2010, \apj, 723, 54

\bibitem[{{Kormendy} \& {Kennicutt}(2004)}]{kk04}
{Kormendy} J., {Kennicutt} Jr. R.~C., 2004, \araa, 42, 603

\bibitem[{{Leitner}(2012)}]{leitner12}
{Leitner} S.~N., 2012, \apj, 745, 149

\bibitem[{{Macci{\`o}} {et~al.}(2012){Macci{\`o}}, {Stinson}, {Brook},
  {Wadsley}, {Couchman}, {Shen}, {Gibson} \& {Quinn}}]{maccio12}
{Macci{\`o}} A.~V., {Stinson} G., {Brook} C.~B., {Wadsley} J., {Couchman}
  H.~M.~P., {Shen} S., {Gibson} B.~K., {Quinn} T., 2012, \apjl, 744, L9

\bibitem[{{Morris} \& {Monaghan}(1997)}]{mm97}
{Morris} J.~P., {Monaghan} J.~J., 1997, Journal of Computational Physics, 136,
  41

\bibitem[{{Moster} {et~al.}(2012){Moster}, {Naab} \& {White}}]{moster12}
{Moster} B.~P., {Naab} T., {White} S.~D.~M., 2012, ArXiv e-prints, 1205.5807

\bibitem[{{Moster} {et~al.}(2010){Moster}, {Somerville}, {Maulbetsch}, {van den
  Bosch}, {Macci{\`o}}, {Naab} \& {Oser}}]{moster10}
{Moster} B.~P., {Somerville} R.~S., {Maulbetsch} C., {van den Bosch} F.~C.,
  {Macci{\`o}} A.~V., {Naab} T., {Oser} L., 2010, \apj, 710, 903

\bibitem[{{Naab} \& {Trujillo}(2006)}]{naab06}
{Naab} T., {Trujillo} I., 2006, \mnras, 369, 625

\bibitem[{{Navarro} {et~al.}(2010){Navarro}, {Ludlow}, {Springel}, {Wang},
  {Vogelsberger}, {White}, {Jenkins}, {Frenk} \& {Helmi}}]{navarro10}
{Navarro} J.~F., {et~al.}, 2010, \mnras, 402, 21

\bibitem[{{Noguchi}(1999)}]{noguchi99}
{Noguchi} M., 1999, \apj, 514, 77

\bibitem[{{Okamoto} {et~al.}(2005){Okamoto}, {Eke}, {Frenk} \&
  {Jenkins}}]{oka05}
{Okamoto} T., {Eke} V.~R., {Frenk} C.~S., {Jenkins} A., 2005, \mnras, 363, 1299

\bibitem[{{Okamoto} \& {Frenk}(2009)}]{of09}
{Okamoto} T., {Frenk} C.~S., 2009, \mnras, 399, L174

\bibitem[{{Okamoto} {et~al.}(2010){Okamoto}, {Frenk}, {Jenkins} \&
  {Theuns}}]{ofjt10}
{Okamoto} T., {Frenk} C.~S., {Jenkins} A., {Theuns} T., 2010, \mnras, 406, 208

\bibitem[{{Okamoto} {et~al.}(2008{\natexlab{a}}){Okamoto}, {Gao} \&
  {Theuns}}]{ogt08}
{Okamoto} T., {Gao} L., {Theuns} T., 2008{\natexlab{a}}, \mnras, 390, 920

\bibitem[{{Okamoto} {et~al.}(2003){Okamoto}, {Jenkins}, {Eke}, {Quilis} \&
  {Frenk}}]{oka03}
{Okamoto} T., {Jenkins} A., {Eke} V.~R., {Quilis} V., {Frenk} C.~S., 2003,
  \mnras, 345, 429

\bibitem[{{Okamoto} {et~al.}(2008{\natexlab{b}}){Okamoto}, {Nemmen} \&
  {Bower}}]{onb08}
{Okamoto} T., {Nemmen} R.~S., {Bower} R.~G., 2008{\natexlab{b}}, \mnras, 385,
  161

\bibitem[{{Oppenheimer} \& {Dav{\'e}}(2006)}]{od06}
{Oppenheimer} B.~D., {Dav{\'e}} R., 2006, \mnras, 373, 1265

\bibitem[{{Parry} {et~al.}(2012){Parry}, {Eke}, {Frenk} \& {Okamoto}}]{parry12}
{Parry} O.~H., {Eke} V.~R., {Frenk} C.~S., {Okamoto} T., 2012, \mnras, 419,
  3304

\bibitem[{{Peebles} \& {Nusser}(2010)}]{peebles10}
{Peebles} P.~J.~E., {Nusser} A., 2010, \nat, 465, 565

\bibitem[{{Pfenniger} \& {Norman}(1990)}]{pfenniger90}
{Pfenniger} D., {Norman} C., 1990, \apj, 363, 391

\bibitem[{{Pohlen} {et~al.}(2000){Pohlen}, {Dettmar}, {L{\"u}tticke} \&
  {Schwarzkopf}}]{pohlen00}
{Pohlen} M., {Dettmar} R.-J., {L{\"u}tticke} R., {Schwarzkopf} U., 2000, \aaps,
  144, 405

\bibitem[{{Price}(2008)}]{price08}
{Price} D.~J., 2008, Journal of Computational Physics, 2271, 10040

\bibitem[{{Robertson} {et~al.}(2006){Robertson}, {Bullock}, {Cox}, {Di Matteo},
  {Hernquist}, {Springel} \& {Yoshida}}]{robertson06}
{Robertson} B., {Bullock} J.~S., {Cox} T.~J., {Di Matteo} T., {Hernquist} L.,
  {Springel} V., {Yoshida} N., 2006, \apj, 645, 986

\bibitem[{{Saitoh} \& {Makino}(2009)}]{sm09}
{Saitoh} T.~R., {Makino} J., 2009, \apjl, 697, L99

\bibitem[{{Sales} {et~al.}(2012){Sales}, {Navarro}, {Theuns}, {Schaye},
  {White}, {Frenk}, {Crain} \& {Dalla Vecchia}}]{sales12}
{Sales} L.~V., {Navarro} J.~F., {Theuns} T., {Schaye} J., {White} S.~D.~M.,
  {Frenk} C.~S., {Crain} R.~A., {Dalla Vecchia} C., 2012, \mnras, 423, 1544

\bibitem[{{Scannapieco} {et~al.}(2012){Scannapieco}, {Wadepuhl}, {Parry},
  {Navarro}, {Jenkins}, {Springel}, {Teyssier}, {Carlson}, {Couchman}, {Crain},
  {Vecchia}, {Frenk}, {Kobayashi}, {Monaco}, {Murante}, {Okamoto}, {Quinn},
  {Schaye}, {Stinson}, {Theuns}, {Wadsley}, {White} \& {Woods}}]{aquila}
{Scannapieco} C., {et~al.}, 2012, \mnras, 423, 1726

\bibitem[{{Scannapieco} {et~al.}(2009){Scannapieco}, {White}, {Springel} \&
  {Tissera}}]{sca09}
{Scannapieco} C., {White} S.~D.~M., {Springel} V., {Tissera} P.~B., 2009,
  \mnras, 396, 696

\bibitem[{{Scannapieco} {et~al.}(2011){Scannapieco}, {White}, {Springel} \&
  {Tissera}}]{scannapieco11}
---, 2011, \mnras, 417, 154

\bibitem[{{Schaye} {et~al.}(2010){Schaye}, {Dalla Vecchia}, {Booth}, {Wiersma},
  {Theuns}, {Haas}, {Bertone}, {Duffy}, {McCarthy} \& {van de Voort}}]{owls}
{Schaye} J., {et~al.}, 2010, \mnras, 402, 1536

\bibitem[{{Shimizu} {et~al.}(2011){Shimizu}, {Yoshida} \&
  {Okamoto}}]{shimizu11}
{Shimizu} I., {Yoshida} N., {Okamoto} T., 2011, \mnras, 418, 2273

\bibitem[{{Shimizu} {et~al.}(2012){Shimizu}, {Yoshida} \&
  {Okamoto}}]{shimizu12}
---, 2012, ArXiv e-prints, 1207.3856

\bibitem[{{Sofue} {et~al.}(2009){Sofue}, {Honma} \& {Omodaka}}]{sofue09}
{Sofue} Y., {Honma} M., {Omodaka} T., 2009, \pasj, 61, 227

\bibitem[{{Spergel} {et~al.}(2003){Spergel}, {Verde}, {Peiris}, {Komatsu},
  {Nolta}, {Bennett}, {Halpern}, {Hinshaw}, {Jarosik}, {Kogut}, {Limon},
  {Meyer}, {Page}, {Tucker}, {Weiland}, {Wollack} \& {Wright}}]{wmap1}
{Spergel} D.~N., {et~al.}, 2003, \apjs, 148, 175

\bibitem[{{Springel} \& {Hernquist}(2003)}]{sh03}
{Springel} V., {Hernquist} L., 2003, \mnras, 339, 289

\bibitem[{{Springel} \& {Hernquist}(2005)}]{sh05}
---, 2005, \apjl, 622, L9

\bibitem[{{Springel} {et~al.}(2008){Springel}, {Wang}, {Vogelsberger},
  {Ludlow}, {Jenkins}, {Helmi}, {Navarro}, {Frenk} \& {White}}]{aquarius}
{Springel} V., {et~al.}, 2008, \mnras, 391, 1685

\bibitem[{{Springel} {et~al.}(2001){Springel}, {White}, {Tormen} \&
  {Kauffmann}}]{spr01}
{Springel} V., {White} S.~D.~M., {Tormen} G., {Kauffmann} G., 2001, \mnras,
  328, 726

\bibitem[{{Stark} {et~al.}(2008){Stark}, {Swinbank}, {Ellis}, {Dye}, {Smail} \&
  {Richard}}]{stark08}
{Stark} D.~P., {Swinbank} A.~M., {Ellis} R.~S., {Dye} S., {Smail} I.~R.,
  {Richard} J., 2008, \nat, 455, 775

\bibitem[{{Wang} {et~al.}(2011){Wang}, {Navarro}, {Frenk}, {White}, {Springel},
  {Jenkins}, {Helmi}, {Ludlow} \& {Vogelsberger}}]{wang11}
{Wang} J., {et~al.}, 2011, \mnras, 413, 1373

\bibitem[{{Weinzirl} {et~al.}(2009){Weinzirl}, {Jogee}, {Khochfar}, {Burkert}
  \& {Kormendy}}]{weinzirl09}
{Weinzirl} T., {Jogee} S., {Khochfar} S., {Burkert} A., {Kormendy} J., 2009,
  \apj, 696, 411

\bibitem[{{Wiersma} {et~al.}(2009){Wiersma}, {Schaye} \& {Smith}}]{wss09}
{Wiersma} R.~P.~C., {Schaye} J., {Smith} B.~D., 2009, \mnras, 393, 99

\end{thebibliography}

\appendix

\section{Properties of the simulated galaxies at redshift 0} \label{appendix:a}

\begin{table*}
\caption{Mass and morphology of the simulated galaxies at redshift 0.  
  The stellar mass, $M_{\rm star}$, and the $i$-band luminosity, $M_i$,  
  of the galaxies are computed by using the stars within the galaxy radius. 
  We present the best fit values of the parameters for 
  the S\'ersic bulges and explanation discs: $\Sigma_{\rm e}$, $R_{\rm e}$ 
  and $n$ in equation~(\ref{eq:sersic}), the disc central surface stellar density, 
  $\Sigma_{\rm d}$, and the disc scale length, $R_{\rm d}$. 
  We also show the $i$-band bulge-to-total luminosity ratio, $(B/T)_i$. 
}
\label{table:gal}
\begin{tabular}{@{}lcccccccc}
\hline
Galaxy 
  & $M_{\rm star}~(M_\odot)$ 
  & $M_i$ (mag) 
  & $\Sigma_{\rm e}~(M_\odot~{\rm kpc}^{-2})$
  & $R_{\rm e}$ (kpc)
  & $n$
  & $\Sigma_0~(M_\odot~{\rm kpc}^{-2})$ 
  & $R_{\rm d}$ (kpc) 
  & $(B/T)_i$ \\
\hline
Aq-C & $4.0 \times 10^{10}$ & -21.3 & $2.2 \times 10^9$ & 0.98 & 1.2 & $1.2 \times 10^8$ & 4.5 & 0.38 \\
Aq-D & $3.1 \times 10^{10}$ & -21.2 & $7.2 \times 10^8$ & 1.1 & 1.4 & $1.1 \times 10^8$ & 5.3 & 0.22 \\
\hline
\end{tabular}
\end{table*}

We here give some of the key properties of our simulated 
galaxies at redshift 0. In Table~\ref{table:gal}, we 
show the mass and $i$-band luminosity of them. We also present 
the best fit values of the parameters in the surface stellar 
density profile fitting. 
For reference, we also show the $i$-band bulge-to-total luminosity 
ratios, $(B/T)_i$, which take lower values than the mass ratios 
because the discs contains more young stars than the bulges. 
The disc scale lengths seem to be too large compared with a 
typical value with observed galaxies of similar types (a few kpc).
Such morphological details are quite sensitive to the implementation 
of star formation and feedback and beyond the scope of this paper. \\

\begin{figure}
  \begin{center}
    \includegraphics[width=8.0cm]{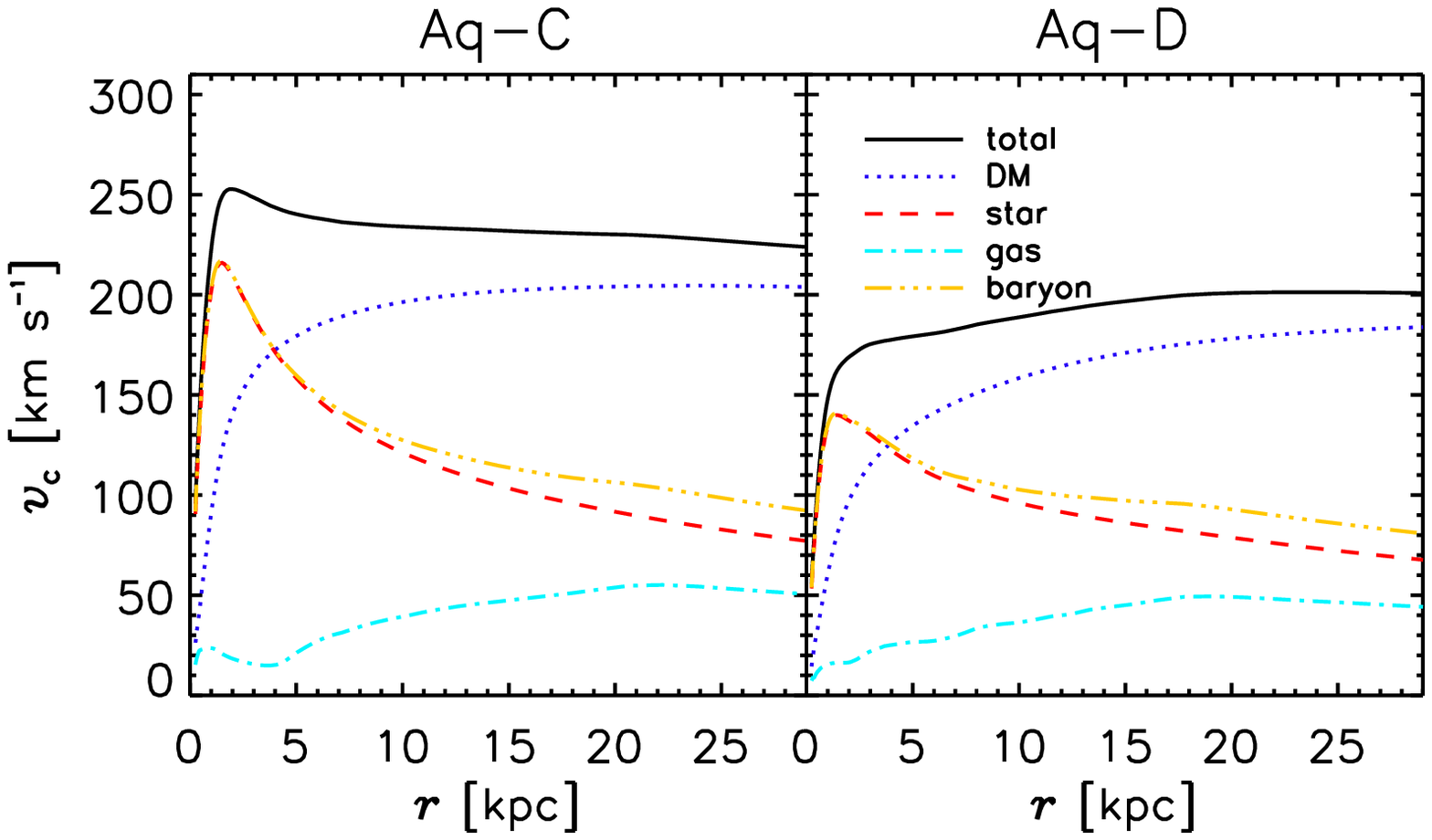}
  \end{center}
  \caption{
    Circular velocity profiles at redshift 0. 
    The circular velocity profiles, $v_{\rm c}(r) = \sqrt{G M(< r)/r}$,  
    are shown by the black solid lines. 
    The blue dotted, red dashed, cyan dot-dashed, and yellow dot-dot-dot-dashed 
    lines respectively indicate the contribution from the dark matter, 
    stellar, gas, and baryonic (stellar plus gas) components. 
  }
  \label{fig:vc}
\end{figure}

In Fig.~\ref{fig:vc}, we show the circular velocity profiles. Note that 
the circular velocity, $v_{\rm c} \equiv \sqrt{G M(<r)/r}$, is different 
from the rotation speed presented 
in Fig.~\ref{fig:velmap}. 
We find that the circular velocity profiles are very flat and the values
are consistent with the Milky Way's value 
\citep[$\sim 200$~km~s$^{-1}$;][see also G3-TO model in \citealt{aquila}]{sofue09}. 
The stellar component dominates within $\sim 4$ kpc. 
Therefore within the bulges, stars dominate in mass and the outer region, 
where the discs exist, is dominated by dark matter.  
\\

\begin{figure}
  \begin{center}
    \includegraphics[width=8.0cm]{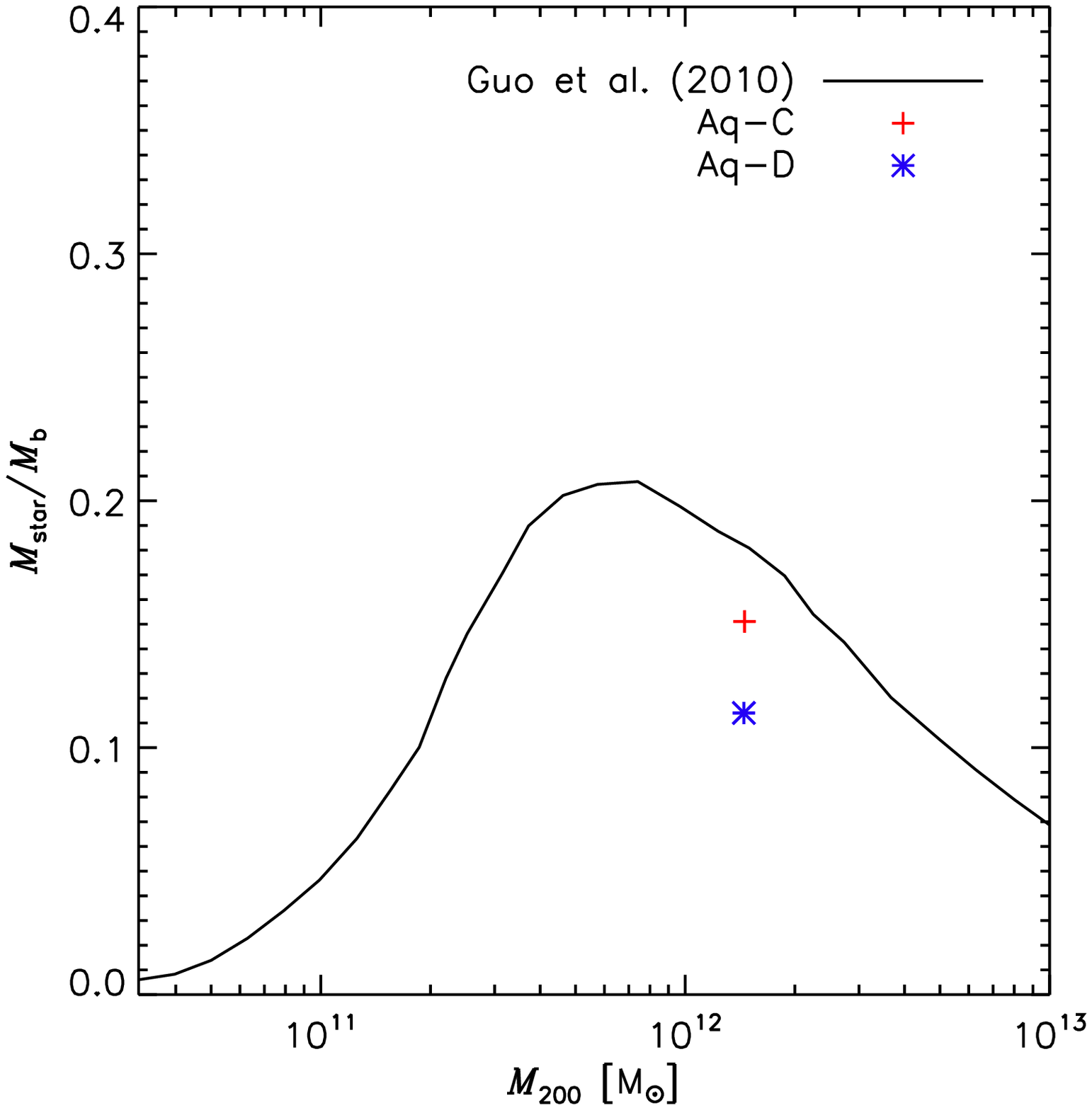}
  \end{center}
  \caption{
    Star formation efficiency as a function of halo mass. 
    The vertical axis shows how many baryons in the halo locked 
    into stars, where 
    $M_{\rm b} \equiv (\Omega_{\rm b}/\Omega_0) M_{200}$ and 
    $M_{\rm 200}$ is the mass of a sphere whose mean density is 200 times the critical density. 
    The black curve is taken from \citet{guo10}, which indicates the values
    required if a $\Lambda$CDM universe is to fit the observed SDSS stellar 
    mass function. 
    The red plus sign and  blue asterisk indicate the values for Aq-C and Aq-D, respectively.  
  }
  \label{fig:bfrac}
\end{figure}

In Fig.~\ref{fig:bfrac}, we show how many baryons in the haloes are locked into 
stars. 
\citet{behroozi10}, \citet{guo10} and  \citet{moster10} have pointed out only 
20\% of baryons in a halo are locked into stars at most. 
\citet{guo10} showed the great majority 
of simulations lock too many baryons into stars \citep[see also][]{aquila}. 
The values for out simulated galaxies are slightly lower than that obtained for the 
SDSS galaxies by \citet{guo10} but they are broadly consistent with their 
result considering the fact that there must be significant halo-to-halo variation. 
The galaxy formation efficiency in Aq-D is lower than that in Aq-C since the 
formation redshift of Aq-D's host halo is more recent and the depth of its 
potential well is shallower; therefore the feedback is more efficient for Aq-D. 
\\

\section{Roles of inflows and outflows} \label{appendix:b}
\begin{figure}
  \begin{center}
    \includegraphics[width=8.0cm]{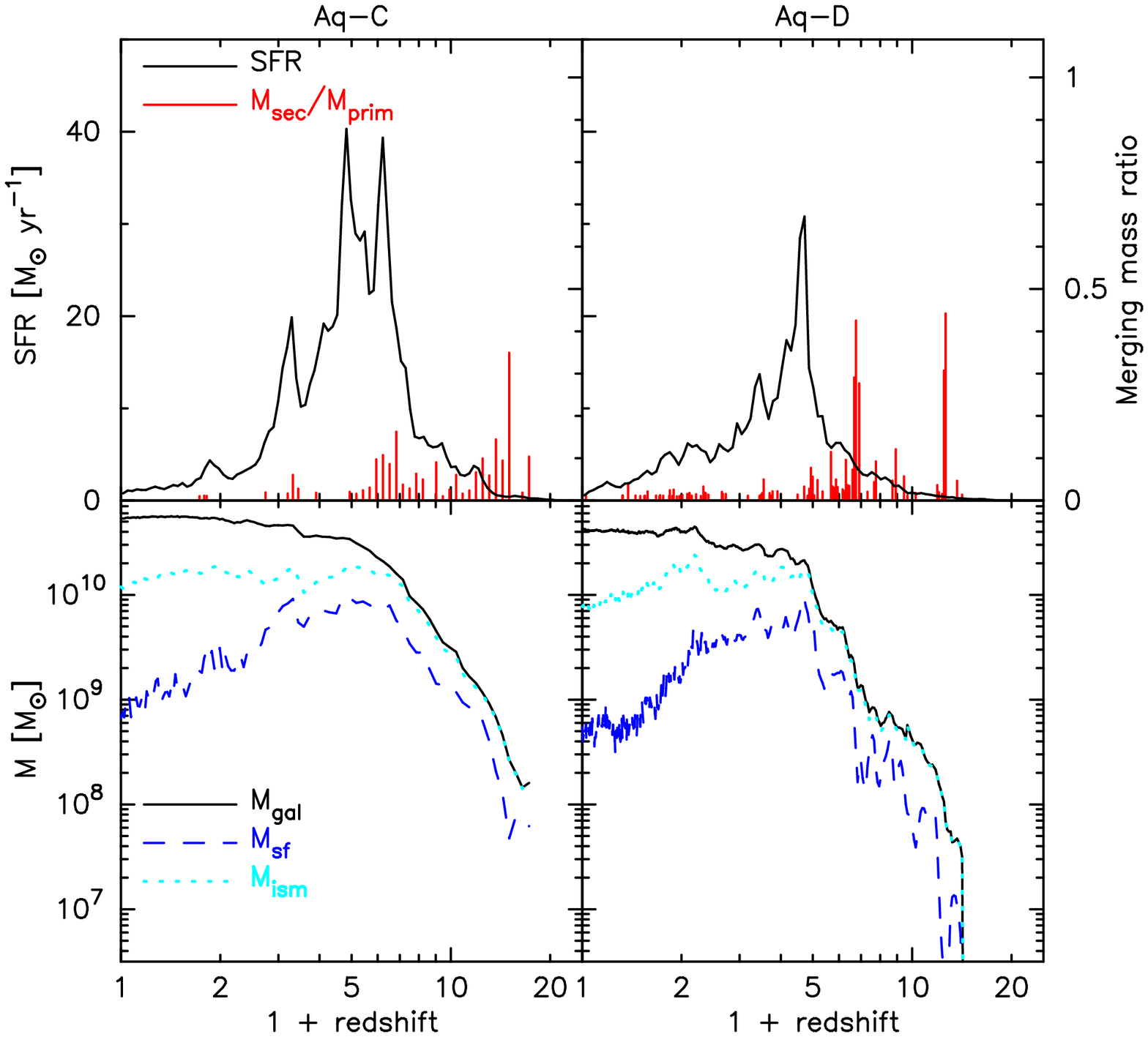}
  \end{center}
  \caption{
    Star formation and merging histories of the simulated galaxies. 
    The left and right panels show evolution of Aq-C and Aq-D, respectively. 
    {\it Upper panels}: Star formation histories of the stars within 
    the galaxy radius ($0.1 r_{\rm vir}$) at redshift 0 are shown by the 
    black lines. The red vertical lines indicate merging mass ratios. 
    All the mergers whose mass ratios are greater than 0.01 are presented. 
    Note that the number of outputs of Aq-D simulation is as four times large 
    as that of Aq-C simulation; hence the number of the vertical lines is 
    larger for Aq-D. 
    {\it Lower panels}: Mass evolution of the main progenitor galaxies.  
    The total (stellar + ISM), star-forming gas, and ISM masses in 
    the main progenitor of each galaxy are shown by the black solid, 
    blue dashed, and cyan dotted lines, respectively.
  } 
  \label{fig:mhistory}
\end{figure}
Here we investigate the roles of inflows and outflows. 
First, we show the star formation and merging histories of the 
simulated galaxies in Fig.~\ref{fig:mhistory}.  
We show the star 
formation histories of stars within the galaxy radius at redshift 0 
and the merging mass ratios. 
We also present the mass evolution of the galaxies and 
the mass in the ISM and star-forming gas in the lower panels. 
The mass evolution indicates that the galaxies are 
quite gas rich until the first starbursts at redshift $\sim 5$ for Aq-C 
and at redshift $\sim 3.5$ for Aq-D since our feedback effectively 
suppresses star formation in small galaxies.  
The major mergers at high redshift are thus {\it wet}.  
\\

Roughly speaking, the starbursts above redshift 2 establish the 
pseudobulges (see Fig.~\ref{fig:bulgeform}) and therefore the quiescent 
star formation below redshift 2 seen in the star formation histories 
of the galaxies as a whole (Fig.~\ref{fig:mhistory}) builds up the discs. 
Since mergers are not correlate with the star formation activity, 
starbursts seem to be triggered by the rapid supply of low 
angular momentum gas. 
Indeed, the galaxy masses are dominated by the ISM when they grow
rapidly as shown in the lower panels of Fig.~\ref{fig:mhistory}. 
This suggests that gas accretion is much faster than star formation. 

\begin{figure}
  \begin{center}
    \includegraphics[width=8.0cm]{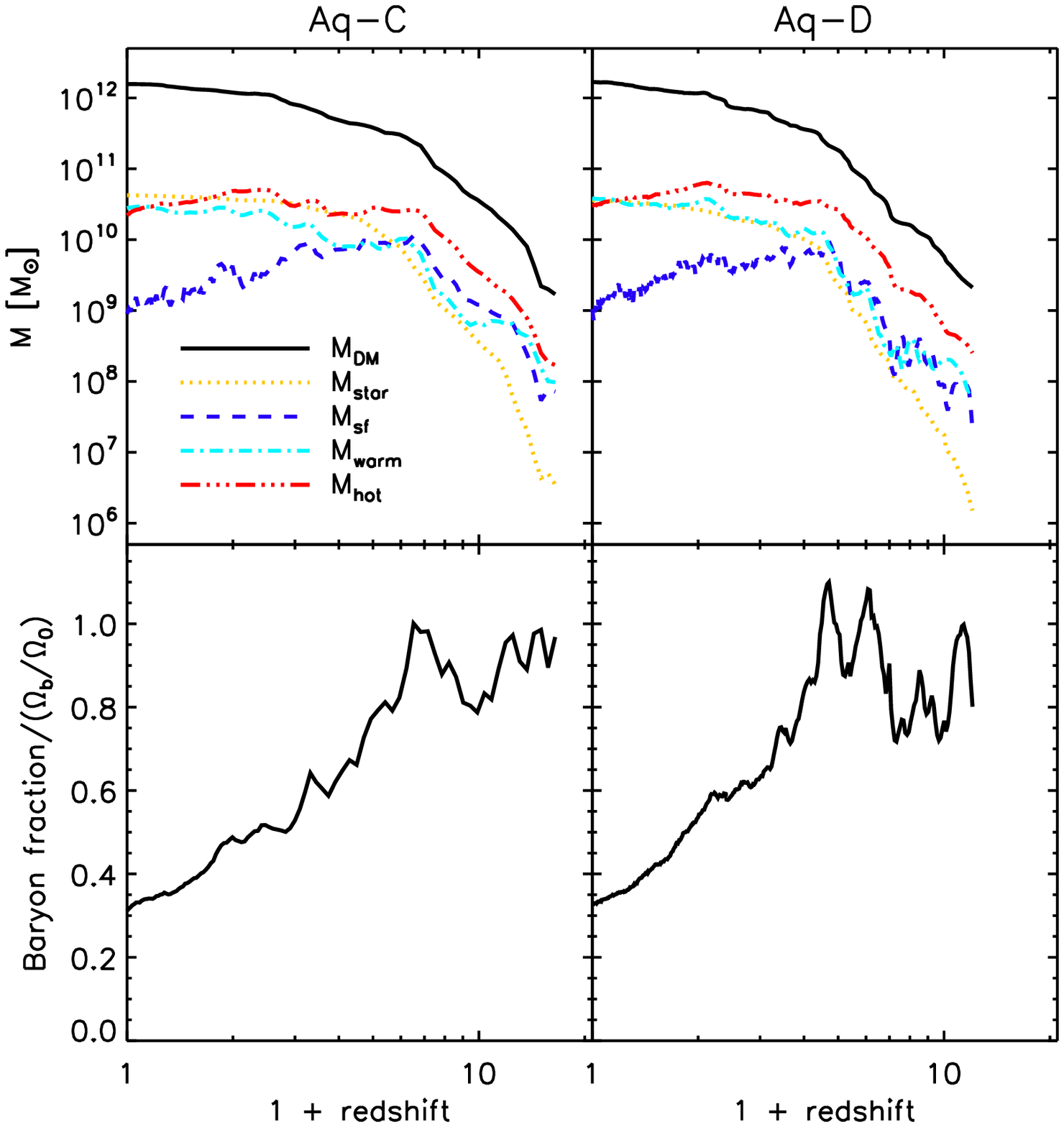}
  \end{center}
  \caption{Mass evolution of the main progenitor haloes. 
    {\it Upper panels}: The mass content in the main progenitor haloes
    as a function of redshift. The black solid, yellow dotted, 
    blue dashed, cyan dot-dashed, and red dot-dot-dot-dashed  
    lines indicate mass in dark matter, stars, star-forming gas, warm gas, 
    and hot gas, respectively. 
    {\it Lower panels}: Evolution of the baryon fractions in the main 
    progenitor haloes. The baryon fractions are normalised by the cosmic mean 
    baryon fraction, $\Omega_{\rm b}/\Omega_0$.  
  }
  \label{fig:massevo}
\end{figure}
Next, we illustrate the mass evolution of the main 
progenitor haloes of Aq-C and Aq-D in Fig.~\ref{fig:massevo}.  
In the upper panels, we show the mass in dark matter, stars, 
star-forming gas, warm gas, and hot gas as functions of redshift, 
where the star-forming gas is high density gas 
($n_{\rm H} > n_{\rm th} = 1.6$~cc$^{-1}$), 
hot gas is that with $T > 10^5$~K or that in the winds, 
and we define the rest as warm gas. \\

Both haloes have virial mass of $\sim 10^{12} \ M_\odot$ 
at redshift 0 as we chose the haloes of the Milky Way's halo mass. 
The amount of the star-forming gas becomes largest at redshift $\sim 5$ 
in Aq-C and at redshift $\sim 3.5$ in Aq-D. Below these 
redshifts the mass of the star-forming gas monotonically decreases 
with decreasing redshift owing to the star formation and feedback. \\  

The effect of winds is depicted in the lower panels of Fig.~\ref{fig:massevo} 
in which we plot the baryon fraction in the main progenitor haloes. 
The baryon fraction is normalised by the cosmic mean value 
$\Omega_{\rm b}/\Omega_0$, 
that is, if a halo has not lost any baryons, the value becomes unity. 
The baryon fraction is high at high redshift and decreases towards lower 
redshift owing to the winds.  
There is an interesting feature between redshift 5--9 in Aq-C and 
3.5--7 in Aq-D, where the baryon fractions increase towards lower 
redshift. 
At these redshifts, the haloes glow rapidly and the gas inflow competes 
the outflow. 
Consequently, the baryon fractions are peaked at redshift $\sim 5$ in 
Aq-C and at redshifts $\sim 3.5$ in Aq-D. 
These redshifts coincide with the redshifts at which the mass of the 
star-forming gas becomes largest; the starbursts occur at these epochs. 
The baryon fractions then sharply drop because strong supernova 
feedback following the starbursts blow out the gas from the haloes.  
During the quiescent star formation (redshift $< 2$), the baryon 
fractions keep decreasing and they are only one third of the 
cosmic mean at redshift 0. 
After the first starbursts, the total baryon mass in the haloes 
are almost constant, while the baryon fractions decreases. 
This means that the mass flux in outflows balances that in the inflows. 
The outflows preferentially remove low angular momentum gas
\citep{gov10, brook11};
on the other hand, low-redshift inflows in general bring in high angular momentum 
gas. 
\\

\begin{figure}
  \begin{center}
    \includegraphics[width=8.0cm]{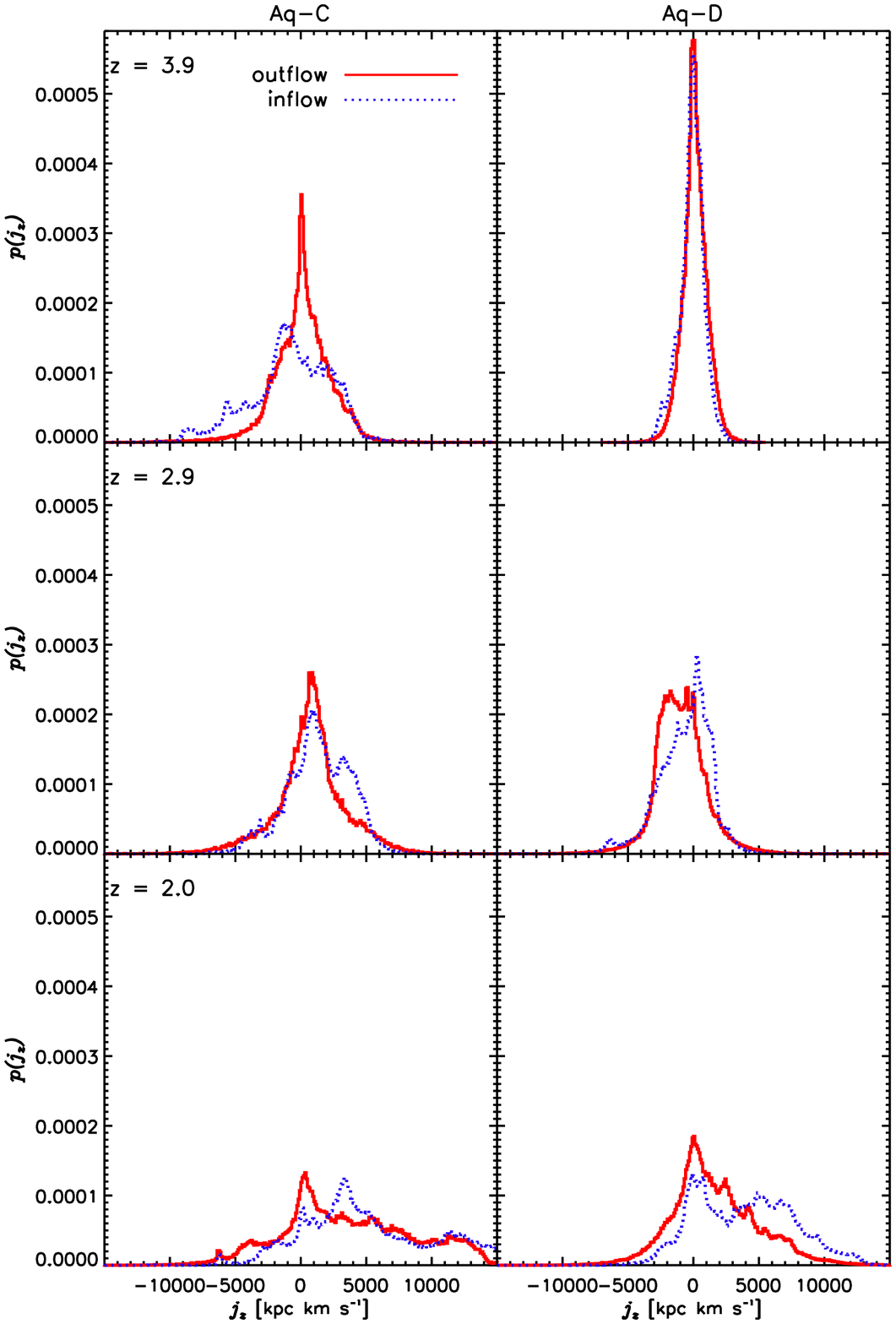}
  \end{center}
  \caption{Distribution of angular momentum of outflows and inflows. 
    Outflows (inflows) are defined as gas moving radially outward (inward) 
    between $0.3 r_{\rm vir}$ and $0.7 r_{\rm vir}$. 
    The red solid and blue dotted lines indicate outflows and inflows, 
    respectively. 
    From top to bottom, the distributions at redshift $\simeq 4$, 3, and 
    2 are shown. 
  }
  \label{fig:flows}
\end{figure}
Fig.~\ref{fig:flows} shows the distribution of angular momentum
of inflows and outflows during the starbursts.  
Outflows (inflows) are defined as gas moving radially outward (inward) 
between $0.3 r_{\rm vir}$ and $0.7 r_{\rm vir}$ where wind particles 
have already been recoupled to the hydrodynamic interactions. 
The $z$-direction is chosen as the direction of the spin of the 
galaxy at a given redshift. 
We find that the angular momentum distributions of the outflows 
are always peaked at $j_z = 0$ showing that the outflows consist 
of low angular momentum gas. 
At redshift $< 3$, the distribution of the angular momentum of 
the inflows show peaks at $j_z > 0$. 
The outflows however remove a large amount of high angular 
momentum gas too. Feedback and star formation prescriptions that
remove low angular momentum gas more efficiently would yield smaller 
bulges \citep[see][]{eris, maccio12}. 

\section{Numerical convergence} \label{appendix:c}

In order to conduct convergence studies, we also simulated the Aq-C halo at two lower 
resolutions with particle masses $\sim 8$ and $\sim 64$ times larger. 
Here we employ the same naming convention as Aquarius, labelling the three runs 
(in order of decreasing resolution) Aq-C-4, Aq-C-5 and Aq-C-6. Aq-C-4 is the one 
we have called `Aq-C'  in the main text of the paper. 
Table~\ref{table:convergence} lists the numerical parameters of each simulation. \\

\begin{table}
  \begin{center}
\caption{
  Numerical parameters employed for the three different resolution simulations: dark matter and 
  SPH particle masses and the gravitational softening length in physical units. The gravitational
  softening length is kept in constant in comoving units until redshift 3 and then fixed in physical 
  units, i.e. the values shown in the table. 
}
\label{table:convergence}
\begin{tabular}{@{}lccc}
\hline
 
  & $M_{\rm DM}~(M_\odot)$ 
  & $M_{\rm SPH}~(M_\odot)$  
  & $\epsilon_{\rm phys}$~(pc) \\
\hline
Aq-C-4 & $2.6 \times 10^{5}$ & $5.8 \times 10^4$ & 257  \\
Aq-C-5 & $2.1 \times 10^{6}$ & $4.7 \times 10^5$ & 514  \\
Aq-C-6 & $1.7 \times 10^{7}$ & $1.7 \times 10^7$ & 1028 \\
\hline
\end{tabular}
\end{center}
\end{table}

The only physical parameter that depends on the numerical resolution is the star formation 
threshold density, $n_{\rm th}$. 
The star formation threshold density for Aq-C-4, -5, and -6 are $n_{\rm th} = 0.1$, 0.4, and 
1.6~cc$^{-1}$, respectively. A factor of 4 higher density is adopted for 
a factor of 8 higher mass resolution. 
This scaling is chosen since, for irradiated nearly isothermal gas with an isothermal 
density profile, halving the gravitational softening length will increase the maximum 
density that is resolved by a facto of 4. \\

Using the same set of the simulations, convergence studies on the satellite population 
was carried out by \citet{parry12} and they found reasonably good convergence of various 
key properties of the simulated satellite galaxies; such as star formation 
histories, mass of the satellites and luminosity functions. 
Comprehensive convergence studies on the central galaxy were presented in \citet{aquila} but 
only Aq-C-5 and Aq-C-6 were used. \\

\begin{figure}
  \begin{center}
    \includegraphics[width=8.0cm]{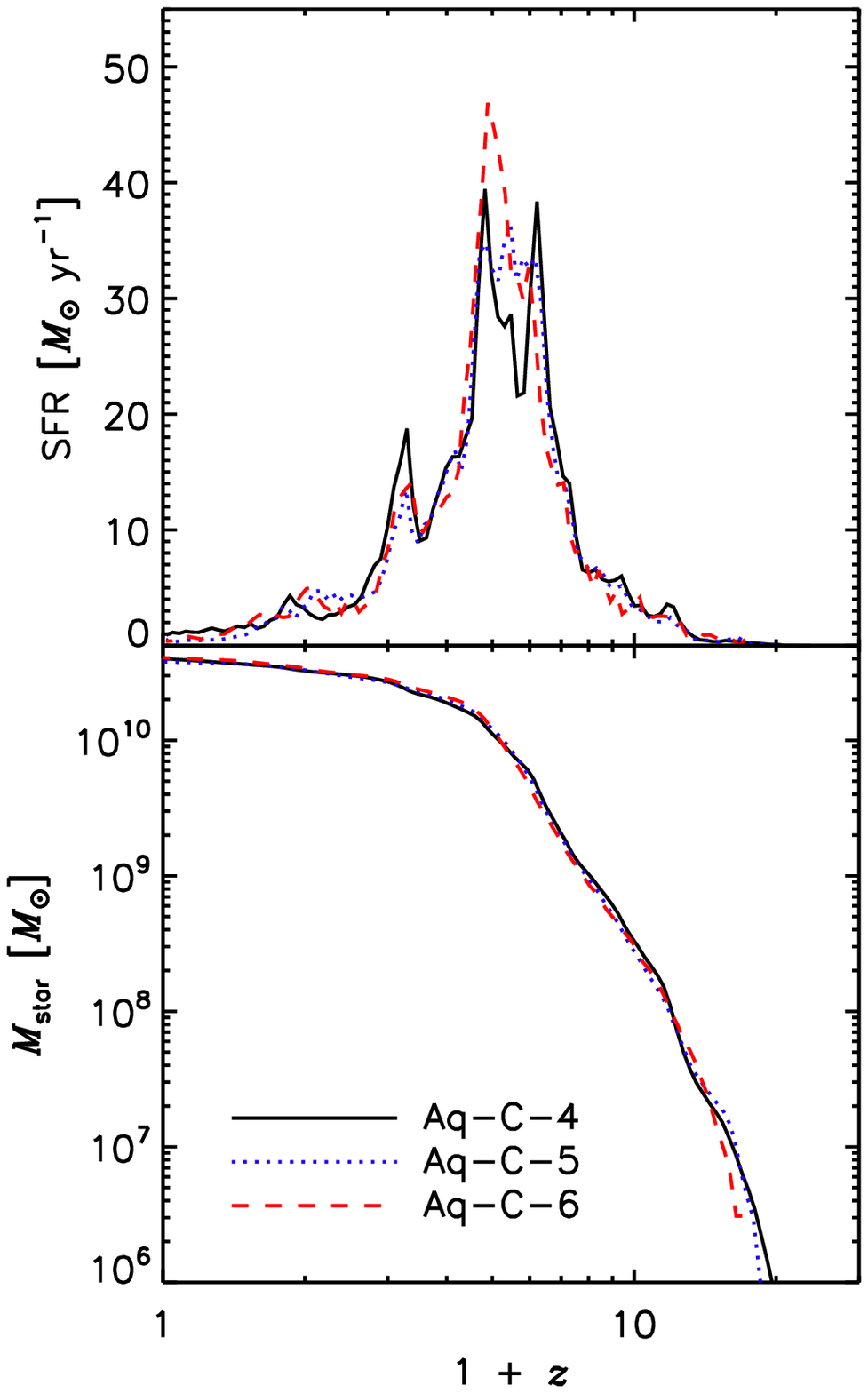}
  \end{center}
  \caption{Formation histories of stars within galaxy radius in three different 
    resolution simulations. 
    The upper panel shows the differential star formation histories while the lower panel
    presents them in cumulative form. 
    The black solid, blue dotted, and red dashed lines corresponds to Aq-C-4, Aq-C-5, and 
    Aq-C-6, respectively. 
  } 
  \label{fig:sfrconv}
\end{figure}

In Fig.~\ref{fig:sfrconv}, we show the star formation histories of these galaxies. 
The results from three different simulations agree quite well, although there are 
some differences in the differential star formation histories during the starbursts. 
Such level of difference will occur even if we only change a seed for a random number 
sequence because of stochasticity of the star formation. \\

\begin{figure}
  \begin{center}
    \includegraphics[width=8.0cm]{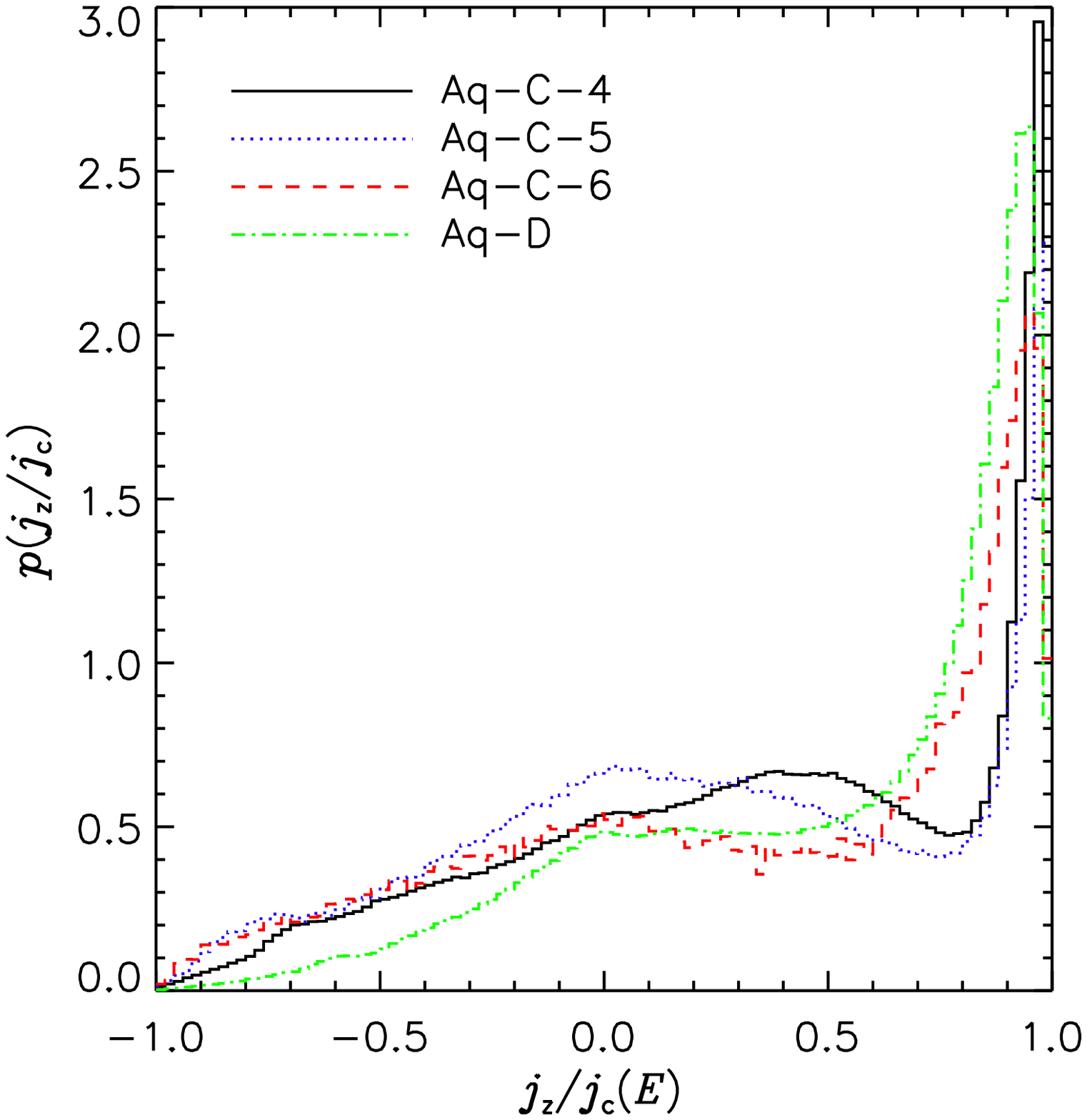}
  \end{center}
  \caption{
    Mass-weighted normalised distribution of orbital circularity, $j_z/j_{\rm c}(E)$, 
    of all stars lying within $r_{\rm vir}$ of the centre, excluding 
    stars identified as members of satellite galaxies.  
    Aq-C-4, Aq-C-5, and Aq-C-6 are indicated by black solid, blue dotted, and red dashed 
    lines, respectively. 
    For reference, we also show the result for Aq-D (corresponding to the level 4 resolution) 
    by the green dot-dashed line. 
    A peak at $j_z/j_{\rm c}(E) \simeq 1$ indicates the existence of a disc component. 
  } 
  \label{fig:orbitconv}
\end{figure}

In order to compare the morphologies of the galaxies, we analyse the 
distribution of the orbital circularity, defined below, of stars in the host 
haloes. We exclude stars belonging to the satellites. 
For each star lying within $r_{\rm vir}$, we compute $j_{\rm z}$, the component 
of specific angular momentum parallel to the $z$-axis. 
We then compute the specific angular momentum, $j_{\rm c}(E)$, of a prograde 
circular orbit with the same binding energy as the particle in question. 
The ratio $j_z/j_{\rm c}(E)$ defines the orbital circularity \citep{aba03b, oka05}. 
The way to compute this quantity is described in \citet{ofjt10} in detail. \\

Fig.~\ref{fig:orbitconv} shows the normalised distributions of the orbital circularity 
of stars within $r_{\rm vir}$ for three difference resolution simulations.  
We also present that for Aq-D for reference. 
A cold disc component has $j_{\rm z}/j_{\rm c}(E) \simeq 1$ and such a component is 
evident in all galaxies. 
There is however no obvious convergence feature. For example, the fraction 
of stars with retrograde orbits (i.e. $j_z < 0$) is largest in Aq-C-5; therefore 
Aq-C-5 is the most spheroid-dominated galaxy of the three\footnote{If we assume the net 
rotation of a spheroidal component is zero as in \citet{aba03b}, 
the mass of a spheroid is given by $2 M(j_z < 0)$.}. 
The position of the second highest peak in Aq-C-4 ($j_z/j_{\rm c} \simeq 0.5$) 
is different from that in lower resolution counterparts, Aq-C-5 and Aq-C-6 
($j_z/j_{\rm c} = 0$). 
Despite the fact that there seems to be no convergence feature, 
it is encouraging that changing resolution does not totally 
change the morphology, that is, they all form well-defined discs. 
By comparing Aq-C-4 with Aq-D, we find Aq-D is more disc dominated than 
Aq-C-4. This is consistent with the result by the decomposition based on 
the surface stellar density profiles. 

\bsp
\label{lastpage}
\end{document}